\documentclass[twocolumn]{aa}
\usepackage{graphicx}
\usepackage{subfigure}
\usepackage{txfonts}
\usepackage{amsmath}
\usepackage{amssymb}
\usepackage{mathtools}
\usepackage{braket}
\usepackage{esint}
\usepackage{bm}
\usepackage[colorlinks=true,allcolors=blue]{hyperref}
\usepackage{float}

\usepackage{comment}
\usepackage{multirow}

\begin{document}
   \title{Cosmological simulations of self-interacting Bose-Einstein condensate dark matter}
   
   \author{S. T. H. Hartman
   \and
   H. A. Winther
   \and
   D. F. Mota
   }

   \institute{Institute of Theoretical Astrophysics, University of Oslo, PO Box 1029, Blindern 0315, Oslo, Norway}
   \date{Received ; accepted }
   
  \abstract
  {Fully 3D cosmological simulations of scalar field dark matter with self-interactions, also known as Bose-Einstein condensate dark matter, are performed using a set of effective hydrodynamic equations. These are derived from the non-linear Schrödinger equation by performing a smoothing operation over scales larger than the de Broglie wavelength, but smaller than the self-interaction Jeans' length. The dynamics on the de Broglie scale become an effective thermal energy in the hydrodynamic approximation, which is assumed to be subdominant in the initial conditions, but become important as structures collapse and the fluid is shock-heated. The halos that form have Navarro-Frenk-White envelopes, while the centers become cored due to the fluid pressures (thermal + self-interaction). The core radii are mostly determined by the self-interaction Jeans' length, even though the effective thermal energy eventually dominates over the self-interaction energy everywhere, a result that is insensitive to the initial smallness of the thermal energy. Scaling relations for the simulated population of halos are compared with Milky Way dwarf spheroidals and nearby galaxies, assuming a Burkert halo profile, and are found to not match, also for core radii $R_c\approx 1\text{kpc}$, which has generally been the value expected to resolve the cusp-core issue. However, the simulations have a limited volume, and therefore a limited halo mass range, include no baryonic physics, and use fiducial cold dark matter initial conditions with a cut-off near the Jeans' length at $z=50$, all of which can affect the halo properties.
  For instance, decreasing the self-interaction strength and moving the cut-off to larger scales show hints of shifting the simulated scaling relations towards the observed trends, which is of particular interest as recent works in the literature suggest this region of parameter space to be in better agreement with large-scale observables and the halo mass function.}
  


\maketitle
   
\section{Introduction}
Unveiling the fundamental nature of dark matter (DM) --- the dominant matter component of our universe --- has been one of the Holy Grails of physics for many decades. Although its particle identity has remained elusive, the collisionless and cold DM (CDM) paradigm, along with a cosmological constant and inflationary initial conditions, has proven extremely successful at explaining a wide range of observables and is known as the $\Lambda$CDM model \citep{Davis1985,Percival2001,Tegmark2004,Trujillo-Gomez2011,Vogelsberger2014,Planck2015,Riess2016, Cyburt2016}. Nevertheless, the quest for a complete theory of DM has spawned numerous expanded models, many of which aim to reproduce the success of CDM while attempting to fix discrepancies between $\Lambda$CDM and observations,
finding physics beyond the standard model of particle physics (SM), or both.

Some of the tensions in $\Lambda$CDM are related to the formation and shape of small-scale structures. CDM, due to its cold and collisionless nature, is very efficient at forming structure at all scales. For instance, $N$-body simulations predict CDM halos to follow the universal Navarro-Frenk-White (NFW) profile \citep{Navarro1996,Navarro1997}, which diverges near the center as $\rho(r)\sim r^{-1}$, whereas measurements instead indicate the central regions of DM profiles of low-mass halos to be flatter, i.e. that low-mass halos are cored rather than cuspy. Furthermore, $\Lambda$CDM predicts a large number of low-mass halos, as well as massive subhalos that should be too big to fail at forming stars, yet these are largely absent in our galactic neighborhood. These issues of $\Lambda$CDM, known as the "cusp-core", "missing-halo", and "too-big-to-fail" problems, respectively, might be due to limitations in accurately modeling baryonic physics and processes that are too small to be resolved in large-scale simulations (see \citet{Weinberg2015,DelPopolo2017,Bullock2017}, and references therein for further discussion on these small-scale discrepancies in $\Lambda$CDM). Alternatively, the solution might lie in the dark sector.

A popular class of DM models beyond CDM involve ultra-light scalar, or pseudo-scalar, particles that have a sufficiently small mass to exhibit wave-like behaviour on astrophysical scales. The free-field case, termed Fuzzy DM (FDM) \citep{Dine1983, Preskill1983,Hu2000,Marsh2016,Hui2017}, with masses as low as $10^{-22}\text{eV}$, have large de Broglie wavelengths $\lambda_{\text{dB}} = 1/mv$ and produce solitonic cores of order $1\text{kpc}$ at the centers of DM halos. Unlike CDM, which clusters on all scales, FDM suppresses small-scale structure because of an effective Jeans' length due to the large de Broglie wavelength. Furthermore, FDM produces interference patterns in its density distribution, which is a very distinct feature of FDM compared to CDM and other DM models, such as warm DM. However, a recent bound on the FDM mass from the Lyman-alpha forest constrains it to $m>2\times 10^{-20}\text{eV}$, and therefore seems to rule out the canonical mass-range generally considered to be needed to solve the small-scale problems of $\Lambda$CDM \citep{Rogers2021}.

Another kind of ultra-light DM are scalar fields with interactions. These can vary in complexity, from single-field DM with self-interactions \citep{Lee1996,Peebles2000,Goodman2000,Arbey2002,Arbey2003,Bohmer2007,Chavanis2011,Rindler-Daller2012}, to multi-field DM with non-trivial couplings that give rise to exotic properties \citep{Matos2000,Bettoni2014,Berezhiani2015,Khoury2016,Ferreira2019}, though a common feature among these models is for the interactions to give rise to a fluid pressure that produces halo cores. The simplest scenario is single-field DM with quartic self-interactions and a negligible de Broglie wavelength, which, along with FDM, is often referred to as Bose-Einstein condensed DM, because they can be regarded as the zero-temperature limit of a boson gas, which in mean-field theory is described by a classical field \citep{Pitaevskii2016}. The term superfluid DM is also frequently used due to the close relationship between Bose-Einstein condensates and superfluidity, though in this work we will instead use "self-interacting Bose-Einstein condensed" (SIBEC) DM, in an effort to emphasize the importance of the self-interactions on the dynamics of the scalar field.

The non-linear structure of SIBEC-DM has largely been investigated near hydrostatic equilibrium, which finds SIBEC-DM halo cores to be independent of the total halo mass and core density. Fitting SIBEC-DM to nearby galaxies reproduces the observed rotation curves for core radii around $1\text{kpc}$ and larger \citep{Zhang2018,Craciun2020}. However, to obtain more realistic SIBEC-DM halo profiles from cosmological structure formation, which are expected transition into the NFW profile outside the cores, one needs to go beyond hydrostatic considerations and use numerical simulations, though some challenges arise in such an endeavor. For instance, the equation of motion for non-relativistic scalar fields, the non-linear Schrödinger equation (NLSE), is very computationally demanding to solve, even in the FDM-limit where the de Broglie wavelength is of astrophysical scale. For SIBEC-DM, where $\lambda_{\text{dB}}$ is much smaller, but still needs to be resolved, the NLSE is even more computationally demanding. Simply setting the terms that describe the dynamics at $\lambda_{\text{dB}}$-scales to zero, which works in the hydrostatic case, does not work in non-linear simulations, as they play an important role in regularizing discontinuities and keeping the numerical solution well-behaved. A scheme that somehow incorporates the de Broglie-scale without actually needing to resolve it is therefore desired, and was recently employed by \citet{Dawoodbhoy2021} and \citet{Shapiro2021}. They used a hydrodynamic formulation of the NLSE that incorporates the de Broglie wavelength as an effective thermal pressure, and performed 1D simulations of collapsing spherical overdensities of SIBEC-DM. They found that in a non-cosmological setting, SIBEC-DM halos with cores of order $1\text{kpc}$ provide a solution to the small-scale issues of the standard model. In fact, they found SIBEC-DM to provide a better solution to both the too-big-to-fail and cusp-core problems than FDM, which by itself struggles to fix these two issues at the same time \citep{Robles2019}. Moving to a cosmological setting, on the other hand, results in an overly large suppression of the formation of small-scale halos for these same core radii. The cores therefore have to be smaller than what is needed to provide a solution to e.g. the core-cusp issue in order to be compatible with the halo mass function. A similar conclusion was drawn by \citet{Hartman2022} using large-scale observables, albeit to a much lesser degree. However, for SIBEC-DM with a CDM-like matter power spectrum at late times \citep{Harko2011,Harko2012,Velteen2012,Freitas2013,Bettoni2014,Freitas2015,Hartman2022}, these constraints are not necessarily true, and cosmological simulations may help test SIBEC-DM in such scenarios.

In this paper we build on the work of \citet{Dawoodbhoy2021} and \citet{Shapiro2021} by implementing the de Broglie-smoothed hydrodynamic formulation of SIBEC-DM into a modified version of the RAMSES code \citep{Teyssier2002}. This code is then used to perform fully 3D cosmological simulations to explore the formation of SIBEC-DM structures and their non-linear dynamics. The paper is structured as follows: In section \ref{sec:basics_SIBEC_DM} the basic theoretical framework for modelling scalar field DM is reviewed, along with the smoothing procedure used to derive the hydrodynamic approximation for SIBEC-DM. In section \ref{sec:numerical_implementation} the numerical aspects of this paper are discussed, including a brief description of how the RAMSES code works, the initial setup of the SIBEC-DM fluid, and a discussion of the limitations of the simulations. Our main results are presented and discussed in section \ref{sec:results}, with final conclusions in section \ref{sec:conclusions}. We use, for the most part, natural units.

\section{Theory of self-interacting Bose-Einstein condensates dark matter}
\label{sec:basics_SIBEC_DM}

Ultra-light scalar DM can be described by a classical field theory, and one of the simplest realizations is the Lagrangian for a complex scalar field $\Psi$ with self-interactions \citep{Li2014},
\begin{equation}
\label{eq:complex_interacting_klein_gordon}
    \mathcal{L} = \frac{1}{2m}g_{\mu\nu}\partial^{\mu}\Psi^{*}\partial^{\nu}\Psi - \frac{1}{2}m|\Psi|^2 - \frac{1}{2}g|\Psi|^4,
\end{equation}
where $m$ is the mass of the scalar field, and $g$ is the coupling parameter for the self-interaction. In the early universe, where DM is much denser and the self-interaction of the scalar field dominates the energy density, the averaged pressure and energy of the field behave as radiation \citep{Matos2001,Li2014}. As the universe expands and becomes more diluted, the interaction energy eventually becomes smaller than the rest energy of the field, and it ceases to be radiation-like. In this non-relativistic phase we can define a new field, $\Psi = \psi e^{-imt}$, whose dynamics are given by the NLSE \citep{Ferreira2020},
\begin{equation}
\label{eq:gross_pitaevskii}
    i\frac{\partial \psi}{\partial t} = \Bigg[\frac{-\nabla^2}{2m} + V\Bigg]\psi,
\end{equation}
where $V = g|\psi|^2 + m\phi$ includes two-body contact interactions, and the gravitational potential $\phi$. The NLSE can be reformulated in a hydrodynamical form by substituting for the wavefunction 
\begin{equation}
\label{eq:madelung_transformation}
    \psi = \sqrt{n} e^{iS} = \sqrt{\frac{\rho}{m}} e^{iS},
\end{equation}
where $n$ is the particle number density, $\rho=mn$ the mass density, and the velocity field is defined by $\bm{v} = \bm{\nabla}S/m$. This gives the Madelung equations \citep{Madelung1926},
\begin{equation}
\label{eq:madelung_continuity}
    \frac{\partial \rho}{\partial t} + \bm{\nabla}\cdot(\rho \bm{v}) = 0,
\end{equation}
\begin{equation}
\label{eq:madelung_momentum}
    \frac{\partial \bm{v}}{\partial t} + (\bm{v}\cdot \bm{\nabla}) \bm{v} + \bm{\nabla}\Bigg(\frac{g\rho}{m^2} - \frac{1}{2m^2}\frac{\nabla^2 \sqrt{\rho}}{\sqrt{\rho}} + \phi \Bigg) = \bm{0},
\end{equation}
which are readily recognizable as an equation for mass conservation, and a variant of the momentum equation. The largest difference from standard hydrodynamics is the absence of an energy equation, the presence of a self-interaction pressure
\begin{equation}
\label{eq:madelung_pressure}
    P_{\text{SI}} = \frac{g\rho^2}{2m^2},
\end{equation}
and a so-called quantum potential
\begin{equation}
\label{eq:madelung_quantum_potential}
    Q = -\frac{1}{2m^2}\frac{\nabla^2 \sqrt{\rho}}{\sqrt{\rho}}.
\end{equation}
The phenomenology of ultra-light DM depends crucially on whether the interaction pressure or the quantum potential dominates. This becomes apparent when considering halo properties of ultra-light DM at equilibrium. In the FDM limit the de Broglie wavelength of the ultra-light DM particles are much larger than the Jeans' length of the self-interaction and is of astrophysical size, such that small-scale structure is stabilized against collapse due to the formation of solitonic cores for masses $m \lesssim 10^{-21}$. The core radii of FDM halos at hydrostatic equilibrium are approximately \citep{Chavanis2011}
\begin{equation}
\label{eq:FDM_halo_radius}
    R_{\text{FDM}} = \frac{10}{GMm^2},
\end{equation}
where $M$ is the total mass of the halo. The opposite limit $Q\ll P_{\text{SI}}/\rho$, which is the case we refer to as SIBEC-DM, but is also known as the Thomas-Fermi approximation, results in the hydrostatic density profile \citep{Chavanis2011}
\begin{equation}
\label{eq:SIBEC_DM_hydrostatic_profile}
    \rho(r) = \rho_0\frac{\sin(Ar)}{Ar},
\end{equation}
where $A = \sqrt{4\pi Gm^2/g}$, and goes to zero at
\begin{equation}
\label{eq:SIBEC_DM_Rc}
    r = R_c = \sqrt{\frac{g\pi}{4Gm^2}}.
\end{equation}
The soliton cores of massive FDM halos are smaller than low-mass halos, a feature confirmed by simulations \citep{Chan2022}, whereas the core radius $R_c$ of halos in the Thomas-Fermi limit is independent of the total halo mass and central density, depending only on the combination $g/m^2$.

Both the NLSE and its hydrodynamic Madelung formulation are widely used to study ultra-light DM across a wide range of scales, from the properties of low-mass halos and galaxies \citep{Chavanis2011,Chavanis2011b,Rindler-Daller2014,Hui2017,Zhang2018,Berezhiani2019,Craciun2020,Lancaster2020} to large-scale simulations \citep{Schive2014,Schive2014b,Schwabe2016,Mocz2017,Mocz2018,Veltmaat2018,Nori2018,Mina2020,Mina2020b,Nori2021,May2021}. However, only the FDM limit has been thoroughly investigated using fully 3D cosmological simulations. No corresponding work has so far been carried out in the Thomas-Fermi limit, in part due to a number of challenges in simulating this kind of system. The hydrodynamic equations obtained by neglecting the quantum potential appear equivalent to regular hydrodynamics with an adiabatic index $\gamma=2$, such that it seems standard numerical schemes should work, but as shock fronts and discontinuities arise the Thomas-Fermi approximation breaks down and the full equations are needed. However, both the NLSE and Madelung equations are computationally demanding to solve due to the high resolution needed to resolve the scales on which the quantum potential operates. This is already a challenge in FDM simulations, but even more so for SIBEC-DM where the characteristic scale of $Q$ is much smaller than the scales of interest given by the interaction pressure, $\lambda_{\text{dB}} \ll R_{\text{c}}$. 

\subsection{Smoothed SIBEC hydrodynamics}
\label{sec:smoothed_SIBEC_DM_hydrodynamics}
An alternative hydrodynamic approximation of SIBEC-DM that includes the effect of the quantum potential without needing to actually resolve $\lambda_{\text{dB}}$ was recently used by \citet{Dawoodbhoy2021} and \citet{Shapiro2021}. It involves using a smoothed phase space representation of the wavefunction, known as the Husimi representation \citep{Husimi1940}, which is obtained by essentially smoothing over $\psi$ with a Gaussian window function of width $\eta$ and Fourier transforming;
\begin{equation}
\label{eq:Husimi_representation}
    \Psi(\bm{x}, \bm{p}, t) = \int\frac{\text{d}^3y}{(2\pi)^{3/2}} \,\frac{e^{-(\bm{x}-\bm{y})^2/2\eta^2}}{(\eta\sqrt{\pi})^{3/2}} \psi(\bm{y},t) e^{-i\bm{p}\cdot(\bm{y}-\bm{x}/2)}.
\end{equation}
Defining the phase space distribution function
\begin{equation}
    \mathcal{F}(\bm{x},\bm{p},t) = |\Psi(\bm{x}, \bm{p}, t)|^2,
\end{equation}
and computing the derivative $\partial\mathcal{F}/\partial t$, using the NLSE $i \partial\psi/\partial t = [-\nabla^2/2m + V]\psi$, and smoothing over scales larger than the de Broglie wavelength $\lambda_{\text{dB}} \ll \eta$, gives an approximate equation of motion that is of the same form as the collisionless Boltzmann equation \citep{Skodje1989,Widrow1993};
\begin{equation}
\label{eq:collisionless_boltzmann_eq}
    \frac{\text{d}\mathcal{F}}{\text{d}t} = \frac{\partial\mathcal{F}}{\partial t} + \frac{p_i}{m}\frac{\partial \mathcal{F}}{\partial x_i} - \frac{\partial V}{\partial x_i}\frac{\partial \mathcal{F}}{\partial p_i} = 0.
\end{equation}
The gravitational potential and the mean-field potential due to self-interactions in $V$ are now given by the smoothed density field, obtained from $\mathcal{F}$ in the same way as from a standard phase space distribution function,
\begin{equation}
    \rho(\bm{x},t) = m\int\text{d}^3 p\,\mathcal{F}(\bm{x},\bm{p},t) = \int\text{d}^3 y\,\frac{e^{-(\bm{x}-\bm{y})^2/\eta^2}}{(\eta\sqrt{\pi})^3}m|\psi(\bm{y},t)|^2.
\end{equation}
Hydrodynamic equations are obtained from the zeroth, first, and second moments of eq. \eqref{eq:collisionless_boltzmann_eq}, which, under the assumption that $\mathcal{F}$ is symmetric and isotropic in $\bm{p}$ around its mean value $\braket{\bm{p}}$ \citep{Veltmaat2018}, are
\begin{equation}
\label{eq:hydro_continuity_eq}
    \frac{\partial \rho}{\partial t} + \bm{\nabla}\cdot\bm{j} = 0,
\end{equation}
\begin{equation}
\label{eq:hydro_euler_eq}
    \frac{\partial \bm{j}}{\partial t} + \bm{\nabla}(P+P_{\text{SI}}) + \rho(\bm{v}\cdot\bm{\nabla})\bm{v} + \bm{v}(\bm{\nabla}\cdot\rho\bm{v}) = -\rho\bm{\nabla}\phi,
\end{equation}
\begin{equation}
\label{eq:hydro_energy_eq}
    \frac{\partial E}{\partial t} + \bm{\nabla}\cdot[(E+P+P_{\text{SI}})\bm{v}] = -\bm{j}\cdot\bm{\nabla}\phi.
\end{equation}
There are two major changes in the above hydrodynamic description of BECs compared to the Madelung equations; first, there is no quantum potential; and second, there is an energy equation in addition to the continuity and momentum equations, with the total energy given by
\begin{equation}
    E = \frac{1}{2}\rho v^2 + U + U_{\text{SI}},
\end{equation}
where $P = \frac{2}{3}U$ and $P_{\text{SI}} = U_{\text{SI}}$. In fact, the small-scale dynamics of the Madelung equations driven by the quantum potential, after smoothing over scales larger than the de Broglie wavelength, behave effectively as an ideal gas with adiabatic index $\gamma=5/3$, with an additional pressure and internal energy due to the self-interactions. The fluid variables, such as the fluid density and momentum, and the effective thermal internal energy and pressure, are defined in the same way as in classical statistical mechanics \citep{Huang1987}, but using the smoothed phase space representation of the wavefunction rather than a phase space distribution function of classical point particles;
\begin{equation}
\label{eq:def_number}
    n(\bm{x},t) = \int\text{d}^3 p\,\mathcal{F}(\bm{x},\bm{p},t),
\end{equation}
\begin{equation}
\label{eq:def_momentum_flux}
    \bm{j}(\bm{x},t) = n\braket{\bm{p}} = \int\text{d}^3 p\,\bm{p}\,\mathcal{F}(\bm{x},\bm{p},t),
\end{equation}
\begin{equation}
\label{eq:def_pressure}
    P(\bm{x},t) = \int\text{d}^3 p\,\frac{|\bm{p}-\braket{\bm{p}}|^2}{3m}\mathcal{F}(\bm{x},\bm{p},t),
\end{equation}
\begin{equation}
\label{eq:def_thermal_energy}
    U(\bm{x},t) = \int\text{d}^3 p\,\frac{|\bm{p}-\braket{\bm{p}}|^2}{2m}\mathcal{F}(\bm{x},\bm{p},t),
\end{equation}
\begin{equation}
\label{eq:def_total_energy}
    E(\bm{x},t) = U_{\text{SI}} + \int\text{d}^3 p\,\frac{p^2}{2m}\mathcal{F}(\bm{x},\bm{p},t).
\end{equation}
In the following we will sometimes refer to the "effective thermal" energy and pressure as simply "thermal".

\subsection{Supercomoving hydrodynamics}
\label{sec:supercomoving_hydrodynamics}
Now that we have a set of fluid equations for SIBEC-DM, we would like to apply these to the growth of structure in a cosmological setting. A convenient set of coordinates for this purpose are the so-called supercomoving coordinates \citep{Martel1998}, which uses the following change in variables in the case of a flat universe dominated by matter and a cosmological constant;
\begin{equation}
\label{eq:supercomoving_transf}
\begin{split}
    & \Tilde{\bm{x}} = \frac{1}{a}\frac{\bm{x}}{L}, \quad 
    \text{d}\Tilde{t} = H_0\frac{\text{d}t}{a^2}, \quad \Tilde{\rho} = a^3\frac{\rho}{\Omega_{m0}\rho_{c0}},\\
    & \Tilde{\bm{u}} = a\frac{\bm{u}}{H_0 L}, \quad \Tilde{P} = a^5\frac{P}{\Omega_{m0}\rho_{c0}H_0^2L^2}, \quad \\
    & \Tilde{m} = mH_0L, \quad \Tilde{g} = g\frac{\Omega_{m0}\rho_{c0}}{a},
\end{split}
\end{equation}
where the peculiar motion $\bm{u} = \bm{v}-H\bm{r}$ is introduced. $L$ is a free length-scale parameter, $\Omega_{m0}$ is the fraction of matter today, and $\rho_{c0}$ the critical energy density today. In these coordinates, the fluid equations in an expanding universe are on a similar form as the standard equations on a static background;
\begin{equation}
\label{eq:supercomoving_hydro_continuity_eq}
    \frac{\partial \Tilde{\rho}}{\partial \Tilde{t}} + \Tilde{\bm{\nabla}}\cdot\Tilde{\bm{j}} = 0,
\end{equation}
\begin{equation}
\label{eq:supercomoving_hydro_euler_eq}
    \frac{\partial \Tilde{\bm{j}}}{\partial \Tilde{t}} + \Tilde{\bm{\nabla}}(\Tilde{P}_{\text{tot}}) + \Tilde{\rho}(\Tilde{\bm{u}}\cdot\Tilde{\bm{\nabla}})\Tilde{\bm{u}} + \Tilde{\bm{u}}(\Tilde{\bm{\nabla}}\cdot\Tilde{\rho}\Tilde{\bm{u}}) = -\Tilde{\rho}\Tilde{\bm{\nabla}}\Tilde{\phi},
\end{equation}
\begin{equation}
\label{eq:supercomoving_hydro_energy_eq}
    \frac{\partial \Tilde{E}}{\partial \Tilde{t}} + \Tilde{\bm{\nabla}}\cdot[(\Tilde{E}+\Tilde{P}_{\text{tot}})\Tilde{\bm{u}}] + \mathcal{H}(3\Tilde{P}_{\text{tot}} - 2\Tilde{U}_{\text{tot}}) = -\Tilde{\bm{j}}\cdot\Tilde{\bm{\nabla}}\Tilde{\phi}.
\end{equation}
where $\Tilde{P}_{\text{tot}} = \Tilde{P} + \Tilde{P}_{\text{SI}}$ and $\Tilde{U}_{\text{tot}} = \Tilde{U} + \Tilde{U}_{\text{SI}}$. The supercomoving Hubble parameter is given by $\mathcal{H} = a^{-1}\text{d}a/\text{d}\Tilde{t}$, the gravitational potential by
\begin{equation}
    \Tilde{\nabla}^2\Tilde{\phi} = \frac{3}{2}a\Omega_m(\Tilde{\rho}-1),
\end{equation}
and expressions for the energies, pressures, and so on, are the same as before, but with supercomoving quantities. A notable difference from regular hydrodynamics is the presence of the Hubble drag in the supercomoving energy equation, which vanishes for the ideal gas pressure with $\gamma=5/3$, but not for the self-interaction pressure. 

\section{Numerical implementation}
\label{sec:numerical_implementation}
A modified version of RAMSES \citep{Teyssier2002} is used to simulate the formation of structure of SIBEC-DM in a cosmological setting. The halo catalogs are obtained using Amiga's Halo Finder (AHF) \citep{Gill2004, Knollmann2009}, and plots of simulation snapshots are made using the python package YT \citep{Turk2010}. RAMSES is a grid based code that uses adaptive mesh refinement (AMR) to focus computational resources through a hierarchy of nested grids, increasing the local resolution in regions of interest. This is done by recursively splitting individual cells into $2^N$ smaller cells, where $N$ is the dimensionality, until a set of refinement criteria are satisfied, doubling the spatial resolution at each new level of refinement. For a box of length $L$, the cell length at refinement level $l$ is $\Delta x^l = L/2^l$.
 
RAMSES was originally designed for cosmological simulations for the formation and evolution of structure at both large and small scales, using an $N$-body particle mesh code for CDM, with which the phase space distribution of DM is sampled through a computationally tractable number of collisionless macroparticles. The gravitational forces acting on these macroparticles are obtained by projecting the DM mass onto the AMR grid and solving the Poisson equation. RAMSES also includes code for studying gas dynamics, using the grid and a Godunov scheme to solve the hydrodynamic equations in supercomoving coordinates. It is this latter part of RAMSES that has been modified to solve the SIBEC-DM hydrodynamic equations, and we use a second-order Godunov scheme with the Rusanov flux (an HLLE Riemann solver) and the van Leer slope limiter to compute the flux at the cell interfaces \citep{Toro2006}. 
 
A number of simulations were run, with different SIBEC-DM interaction strengths (given by the core radius $R_c$), initial conditions, and box sizes to balance reaching sufficient spatial resolution to resolve the cores of the emerging SIBEC-DM halos, while also forming enough such halos to study their properties. Initial conditions were generated with the MUSIC code \citep{Hahn2011} using the standard CDM matter power spectrum, but with a cut-off in power below a certain scale to include the effect of a non-zero sound speed,
\begin{equation}
    P_{\text{SIBEC}}(k) = f_{\text{cut}}(k) P_{\text{CDM}}(k).
\end{equation}
$P_{\text{CDM}}$ is the standard CDM matter power spectrum, and $f_{\text{cut}}$ is a cut-off factor, given by the Heaviside function $f_{\text{cut}}(k) = \Theta(k_{\text{cut}}-k)$, where $k_{\text{cut}}$ is the cut-off scale. We have taken this scale to be the Jeans' length of SIBEC-DM at the time the simulation starts,
\begin{equation}
    k_{\text{cut}} = k_{\text{J}} = \frac{\pi}{R_c}.
\end{equation}
An alternative to the sharp cut-off above is a smooth cut-off, and a comparison with output from the Boltzmann code CLASS, modified to include SIBEC-DM \citep{Hartman2022}, shows that the function
\begin{equation}
\label{eq:smooth_f_cutoff}
    f_{\text{cut}}(k) = e^{-(k/k_{\text{cut}})^3},
\end{equation}
with $k_{\text{cut}}$ given by the effective sound horizon $r_s$ of SIBEC-DM,
\begin{equation}
    r_{\text{s}} = \int_{0}^{\tau_{\text{init}}} \text{d}\tau \,c_s = \int^{\infty}_{z_{\text{init}}} \text{d}z \frac{c_s}{H}, 
\end{equation}
provides a good fit to the suppression of power in a SIBEC-DM universe relative to $\Lambda$CDM. The SIBEC-DM sound speed $c_s$, which includes the early radiation-like epoch during which the interaction energy dominates, is approximately given by \citep{Hartman2022}
\begin{equation}
    \label{eq:SIBEC_cs2}
    c_{s}^2 = \frac{\partial \bar{P}}{\partial \bar{\rho}} = \frac{1}{3}\frac{1}{1 + a^3/6\omega_{0}},
\end{equation}
where $\omega_0$ is the SIBEC-DM equation of state today,
\begin{equation}
\label{eq:w0_limit_hydrostatic_equilibrium}
    \omega_0 = \frac{\bar{P}_0}{\bar{\rho}_0} = \frac{g\bar{\rho}_0}{2m^2} \approx 10^{-15} \left(\frac{R_c}{1\text{kpc}}\right)^2.
\end{equation}
The resulting cut-off scale is
\begin{equation}
\label{eq:k_cutoff_smooth}
    k_{\text{cut}} = \frac{1}{2.2}k_{\text{s}} = \frac{1}{2.2}\frac{2\pi}{r_s} \approx 0.3 h\text{Mpc}^{-1} \Bigg(\frac{R_c}{1\text{kpc}}\Bigg)^{-2/3},
\end{equation}
This form for $f_{\text{cut}}$ provides more realistic initial conditions for SIBEC-DM as described by the Lagrangian eq. \eqref{eq:complex_interacting_klein_gordon}, but the sharp cut-off near the Jeans' scale at the start of the simulation has the advantage of providing more initial power, such that a larger number of SIBEC-DM halos form inside the simulation box and can be used in our analysis. In fact, we can estimate the halo masses that are affected by the cut-off given by eq. \eqref{eq:k_cutoff_smooth},
\begin{equation}
\label{eq:M_cut}
    M_{\text{cut}} \approx \frac{4\pi}{3}\rho_{\text{dm},0}\Bigg(\frac{\pi}{k_{\text{cut}}}\Bigg)^3 \approx 5\times 10^{14}M_{\odot} \, \Bigg(\frac{R_c}{1\text{kpc}}\Bigg)^2.
\end{equation}
This poses a serious challenge for SIBEC-DM, as pointed out by \citet{Shapiro2021}. Self-interacting scalar field DM with core radii of order $1\text{kpc}$ strongly suppresses the formation of halos with masses below $M_{\text{cut}}\approx 5\times 10^{14}M_{\odot}$, making SIBEC-DM with these initial conditions an unfavorable candidate for explaining the cusp-core problem\footnote{The cut-off given by eq. \eqref{eq:k_cutoff_smooth} that we found in an earlier work \citep{Hartman2022} differs slightly from what \citet{Shapiro2021} obtained, but the conclusions regarding the over-suppression of halos remain the same.}. Furthermore, from a numerical point of view, the large difference between the cut-off scale $\lambda_{\text{cut}} = \pi/k_{\text{cut}}$ and the SIBEC core radius $R_c$, i.e. $\lambda_{\text{cut}} \gg R_c$, also poses a challenge, since it requires a both a large simulation box and very high spatial resolution to simultaneously investigate the formation of SIBEC-DM structure and resolve the SIBEC-DM halo cores. For instance, $R_c=1\text{kpc}$ gives the cut-off length scale $\lambda_{\text{cut}}\approx 15\text{Mpc}$, while $R_c=10\text{pc}$ (which \citet{Shapiro2021} finds as an upper limit to avoid overly suppressing the formation of small-scale halos) gives $\lambda_{\text{cut}}\approx 0.7\text{Mpc}$. We therefore do not run simulations with the initial conditions given by eqs. \eqref{eq:smooth_f_cutoff} and \eqref{eq:k_cutoff_smooth}. Nevertheless, the present work may provide an indication for how SIBEC-DM generally behaves in a cosmological setting, and will certainly help us better understand the properties of halos with $R_c \sim 1\text{kpc}$ and their populations in scenarios that assume a matter power spectrum close to CDM at late times, and are therefore not limited by eq. \eqref{eq:M_cut} \citep{Harko2011,Harko2012,Velteen2012,Freitas2013,Bettoni2014,Freitas2015,Hartman2022}.

Initial conditions for the effective thermal energy of SIBEC-DM present in eqs. \eqref{eq:supercomoving_hydro_continuity_eq}, \eqref{eq:supercomoving_hydro_euler_eq}, and \eqref{eq:supercomoving_hydro_energy_eq} must also be provided and is in principle given by the underlying wavefunction, although in the hydrodynamic approximation it has been integrated away with the smoothing kernel in eq. \eqref{eq:Husimi_representation}. We therefore make an ansatz that the effective thermal pressure $P$ is small compared to the interaction pressure $P_{\text{SI}}$ at the start of the simulation, and set $\Tilde{P}(z_{\text{init}})=\zeta \Tilde{P}_{\text{SI}}(z_{\text{init}})$ with $\zeta\ll 1$. The supercomoving effective thermal pressure at the background level is constant in time, while the interaction pressure decreases as $\Tilde{P}_{\text{SI}} \propto a^{-1}$, so $\zeta$ should be sufficiently small that $\Tilde{P}$ is largely negligible until the first structures start to form, at which point the effective thermal energy increases rapidly as matter is accreted onto halos and is heated.

\begin{figure}
    \centering
    \includegraphics[width=0.98\linewidth]{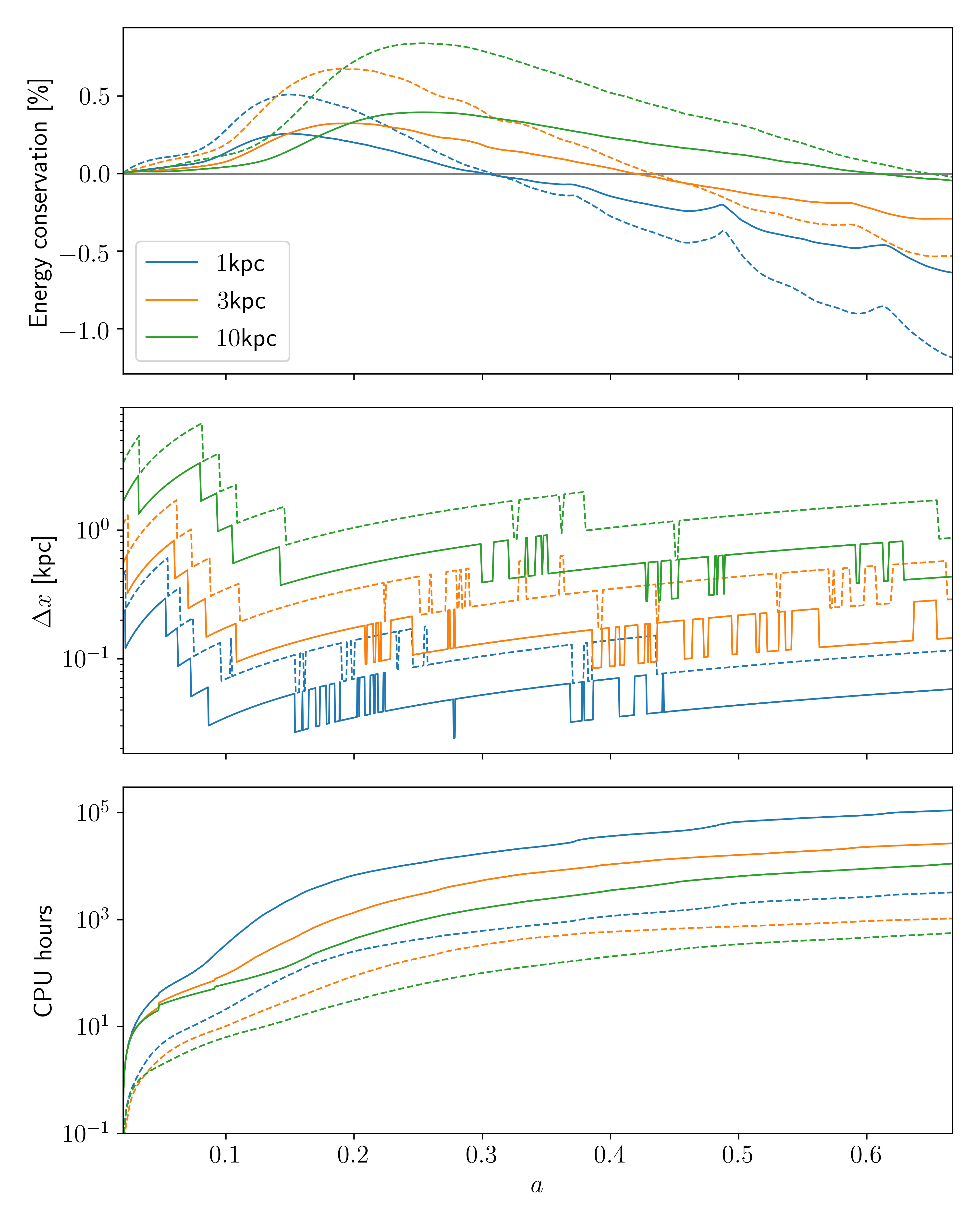}

    \caption{Energy conservation (\textit{top}), the smallest physical cell length (\textit{middle}), and the runtime in CPU hours (\textit{bottom}) for the main suite of simulations. Runs with a minimum refinement level of 8 are shown in solid lines, and 7 in dotted lines.}
\label{fig:simulation_summaries}
\end{figure}

\begin{figure}
    \centering
    \includegraphics[width=0.98\linewidth]{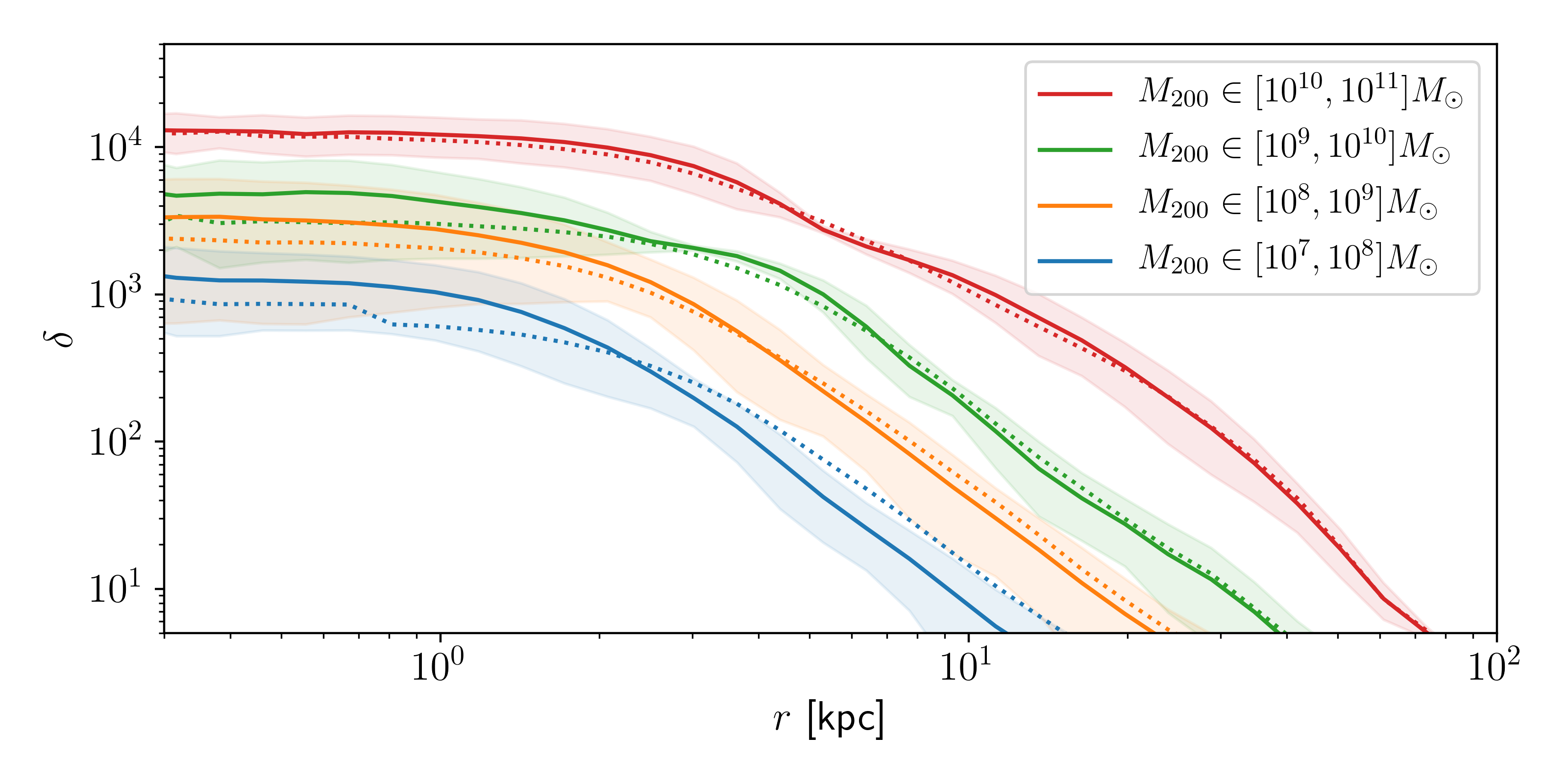}
    \includegraphics[width=0.98\linewidth]{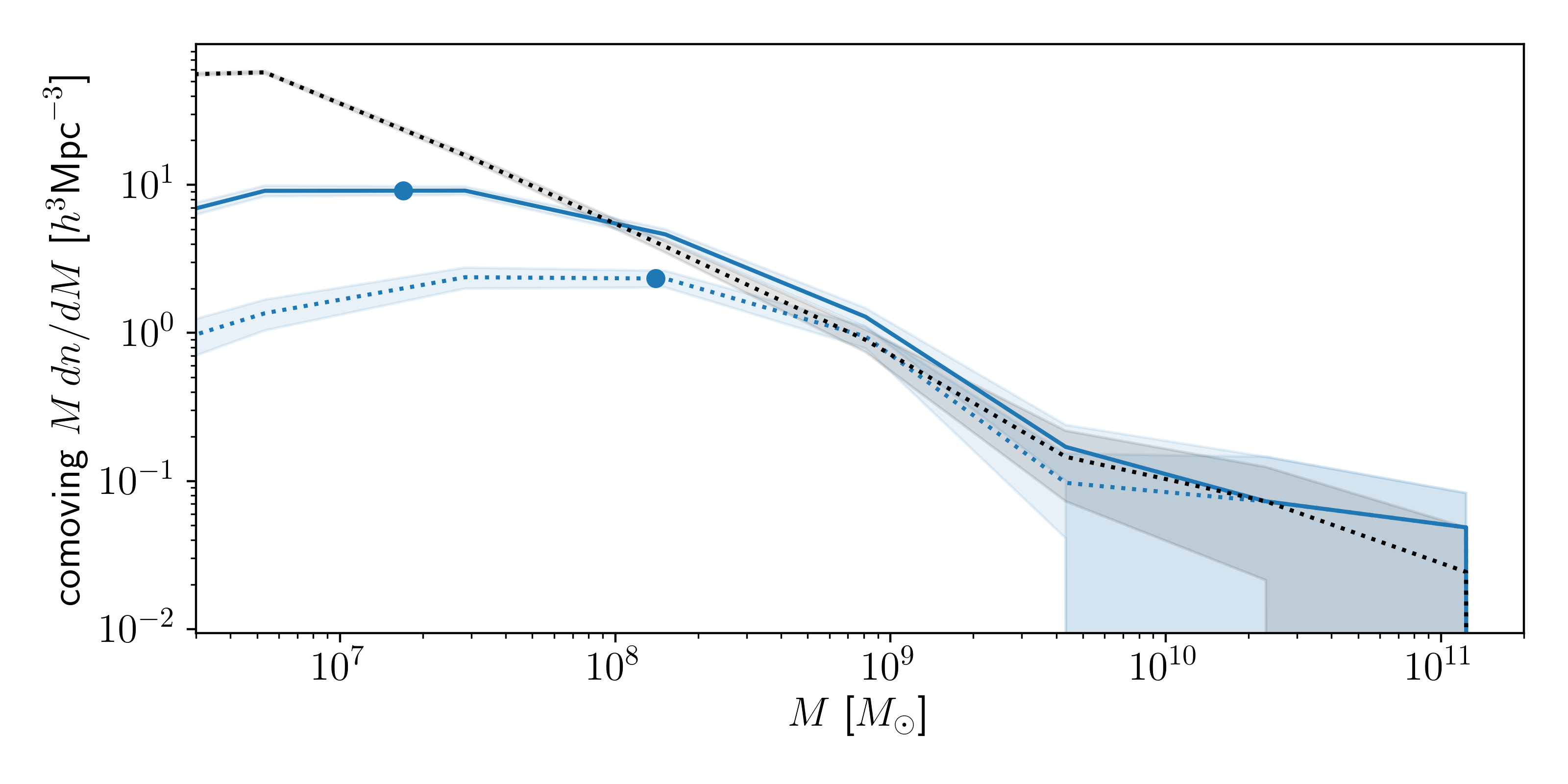}

    \caption{Binned SIBEC-DM halo profiles and the HMF for $R_c=1\text{kpc}$ at $z=1.5$, with minimum refinement level 7 (\textit{dotted}) and 8 (\textit{solid}). The HMF of CDM for the same box size and initial resolution is included in black. The halo mass limit $M_{\text{min}}$ are indicated by dots. The standard deviation from the binned halo profile mean for level 8, as well as the Poisson error in the HMF, are shown in shaded.}
\label{fig:simulation_profiles_and_HMF_refinement_comparison}
\end{figure}

The simulations performed in this work are summarized in table \ref{tab:simulation_overview}, and an overview of the conservation of energy, resolution, and runtime for a few of these are shown in figure~\ref{fig:simulation_summaries} at two different initial refinement levels. The conservation of energy is satisfied to within $1\%$, and the Jean's length is resolved in over-dense regions with the use of AMR at all times, more so at later times as the initial perturbations collapse to form halos. On the other hand, the initial resolution of the grid, which is at the lowest level of refinement, places a lower bound on the halo masses that are accurately evolved. An estimate of this bound is the total mass enclosed by a minimum number of cells $N_{\text{min}}$ on the initial grid,
\begin{equation}
    M_{\text{min}} = \rho_{\text{dm,0}}\,N_{\text{min}}\Bigg(\frac{L}{2^{l_{\text{init}}}}\Bigg)^3.
\end{equation}
We find that $N_{\text{min}}=300$, i.e. the mass enclosed by an approximately $7\times 7\times 7$ cube, provides a reasonable estimate, as shown in figure~\ref{fig:simulation_profiles_and_HMF_refinement_comparison}. In this case $M_{\text{min}}\approx 10^{7}M_{\odot}$ for $l_{\text{init}}=8$, whereas at $l_{\text{init}}=7$ the mass limit is $M_{\text{min}}\approx 10^{8}M_{\odot}$. The profiles are largely the same except in the lowest mass bin, which is below $M_{\text{min}}$ for $l_{\text{init}}=7$. Additionally, the halo mass function (HMF) is suppressed below $M_{\text{min}}$, hence halos with $M_{200}<M_{\text{min}}$ are discarded in the following analysis. It should be noted that the higher resolution yields slightly denser halos even for the masses that are most resolved, and that the HMF above $M_{\text{min}}$ is not exactly identical at the two refinement levels, as one would expect. This indicates our simulations would benefit from further grid refinement, in particular the Rc1 run, which has the highest ratio of both $M_{\text{min}}/M_{\text{cut}}$ and $\Delta x_{\text{min}}/R_c$. Unfortunately, the computational cost of further increasing the resolution was prohibitive, since at $l_\text{init}=8$ the simulation used up to a hundred thousand CPU hours to reach $z=0.5$, while an increase to the next refinement level $l_\text{init}=9$ is expected to use a few million CPU hours.

\bgroup 
\def\arraystretch{1.3}
\begin{table*}[t]
\caption{An overview of the simulations run. All were run with initial/minimum refinement level 8 (7) and maximum level 15 (14) from $z=50$ to $z=0.5$. Unless stated otherwise, when simulation runs are referred to by only $1$kpc, $3$kpc , or $10$kpc, it means Rc1, Rc3, and Rc10 with minimum refinement level 8.\newline}              
\label{tab:simulation_overview}      
\centering                                      
\begin{tabular}{l l l l l l l}          
\hline\hline
Label & $R_c$ [$\text{kpc}$] & Cut-off and $k_{\text{cut}}$ & Initial $\zeta=P/P_{\text{SI}}$ & Boxsize $L$ [$\text{Mpc}/h$] & $M_{\text{min}}$ $[M_{\odot}]$ & $M_{\text{cut}}$ $[M_{\odot}]$ \\
\hline
Rc1 & 1                    & $k_{\text{J}}$          & $10^{-2}$                 & 2 & $1.7\times 10^7$ & $7.2\times 10^6$ \\
Rc3 & 3                    & $k_{\text{J}}$          & $10^{-2}$                 & 5 & $2.6\times 10^8$ & $1.9\times 10^8$ \\
Rc3-b & 3                    & $k_{\text{J}}/2$          & $10^{-2}$                 & 5 & $2.6\times 10^8$ & $1.6\times 10^9$ \\
Rc3-c & 3                    & $k_{\text{J}}$          & $0.1$                 & 5 & $2.6\times 10^8$ & $1.9\times 10^8$ \\
Rc3-d & 3                    & $k_{\text{J}}$        & $0.5$                 & 5 & $2.6\times 10^8$ & $1.9\times 10^9$ \\
Rc10 & 10                   & $k_{\text{J}}$          & $10^{-2}$                 & 15 & $7.1\times 10^9$ & $7.2\times 10^{9}$ \\
\hline\hline
\end{tabular}
\end{table*}
\egroup 

\section{Results}
\label{sec:results}
\begin{figure}
    \centering
    \subfigure{\includegraphics[width=0.98\linewidth]{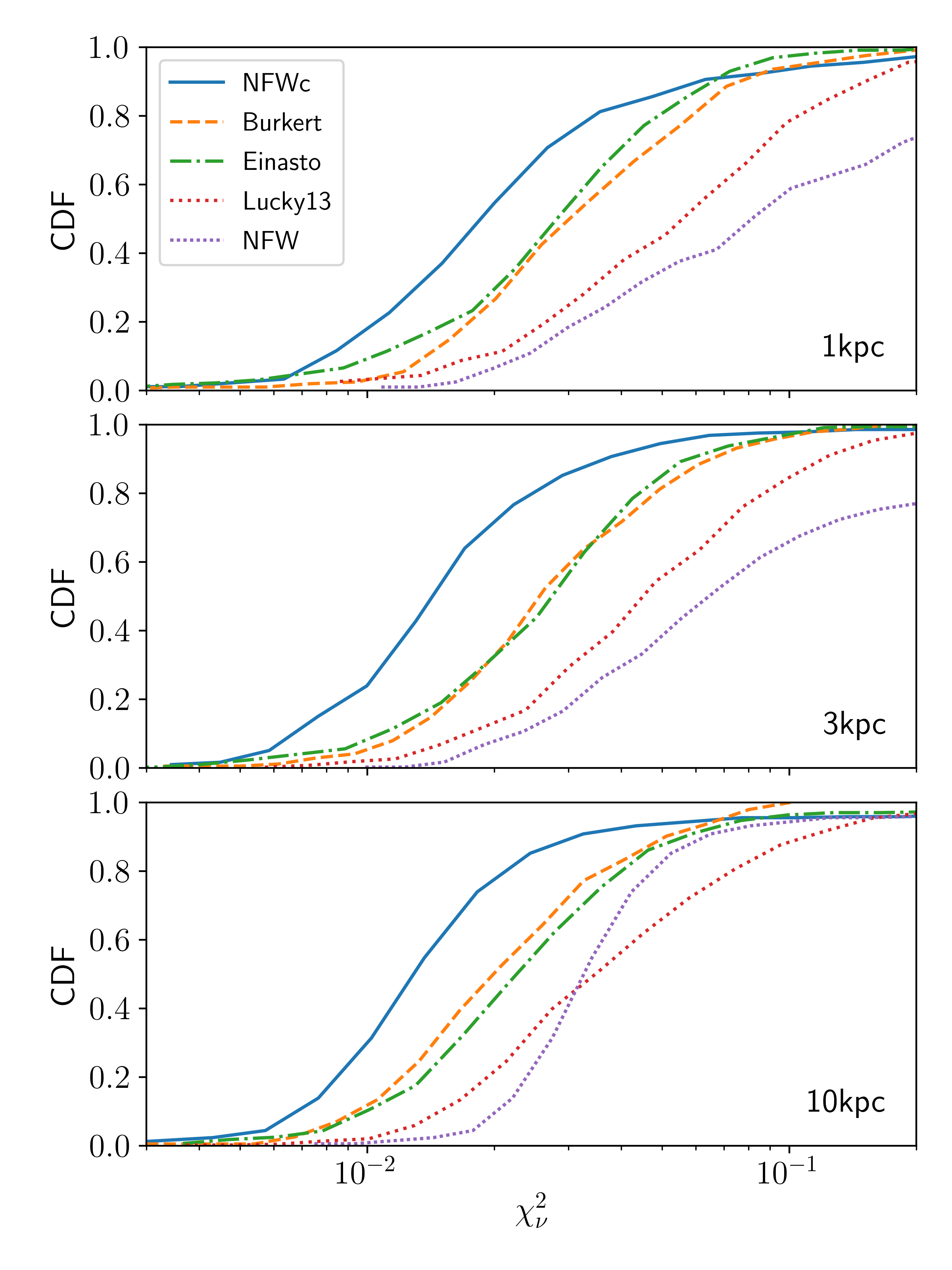}}

    \caption{Cumulative distribution functions for different cored halo profiles from SIBEC-DM simulations with $R_c=1\text{kpc}$ (\textit{upper}), $R_c=3\text{kpc}$ (\textit{middle}), and $R_c=10\text{kpc}$ (\textit{lower}) at $z=0.5$. This shows the fraction of the halos that has $\chi_v^2$ smaller than a given value, and therefore how well the fitting function generally describes the halos in our simulation.}
\label{fig:fitting_CDF_comparisons}
\end{figure}

In the cosmological simulations the SIBEC-DM collapses and forms halos with a cored inner structure and a CDM-like envelope, which we find to be well-fitted by a cored NFW (NFWc) profile,
\begin{equation}
\label{eq:NFWc_profile}
    \delta_{\text{NFWc}}(r) = \Big[\delta_c^{-1} + \delta_{\text{NFW}}^{-1}\Big]^{-1}.
\end{equation}
The profile is given in terms of over-density $\delta = \rho/\bar{\rho}$, where $\delta_c$ is the central over-density, and $\delta_{\text{NFW}}$ is the NFW profile,
\begin{equation}
\label{eq:NFW_profile}
    \delta_{\text{NFW}}(r) = \frac{\delta_s}{\frac{r}{r_s}\Big(1 + \frac{r}{r_s}\Big)^2}.
\end{equation}
We define the core radius $r_c$ as where the density has dropped to $50\%$ of its central value, $\delta(r_c) = 0.5\delta(0)$. At hydrostatic equilibrium we would have $r_c \approx 0.6R_c$.
A number of other cored profiles could also be used to fit the SIBEC-DM halos. In \citet{Li2020} several DM profiles were fitted to the Spitzer Photometry \& Accurate Rotation Curves (SPARC) dataset of rotation curves of nearby galaxies \citep{Lelli2016}, such as the Burkert profile \citep{Burkert1995},
\begin{equation}
\label{eq:Burkert_profile}
    \delta_{\text{Burkert}}(r) = \frac{\delta_c}{\Big(1+\frac{r}{r_s}\Big)\Big[1 + \Big(\frac{r}{r_s}\Big)^2\Big]},
\end{equation}
the Einasto profile \citep{Einasto1965},
\begin{equation}
\label{eq:Einasto_profile}
    \delta_{\text{Einasto}}(r) = \delta_s\exp\Bigg\{-\frac{2}{\alpha_{\epsilon}}\Bigg[\Bigg(\frac{r}{r_s}\Bigg)^{\alpha_{\epsilon}} - 1\Bigg]\Bigg\}
\end{equation}
and the so-called Lucky13 profile \citep{Li2020},
\begin{equation}
\label{eq:Lucky13_profile}
    \delta_{13}(r) = \frac{\delta_c}{\Big[1 + \Big(\frac{r}{r_s}\Big)^3\Big]}.
\end{equation}
These are fitted to the SIBEC-DM halos by minimizing the cost function
\begin{equation}
    \chi_{\nu}^2 = \frac{1}{N_d-N_f}\sum_{i=1}^{N_d}\frac{\big[\delta_i - \delta(r_i)\big]^{2}}{\delta_i^2},
\end{equation}
with respect to the $N_f$ free parameters of the density function $\delta(r)$, where $N_d$ is the number of radial points $\delta_i$ from the simulation that we are fitting the density profile to. Each halo in the simulation gives a value for $\chi_{\nu}^2$, and the corresponding cumulative distribution function (CDF) for the set of these tells us what fraction of the halos has $\chi_{\nu}^2$ smaller than a given value, and therefore how well the fitting function generally describes the halos. In figure~\ref{fig:fitting_CDF_comparisons} the CDF for the different types of profiles are shown, from which we find that the NFWc profile provides the best fit overall. Of the cored profiles used by \citet{Li2020} to fit the DM halos in the SPARC dataset, the Burkert and Einasto profiles are best suited and perform about the same, but as Burkert is the simpler of the two, we will use it in the following.

\begin{figure}
    \centering
    \subfigure{\includegraphics[width=0.98\linewidth]{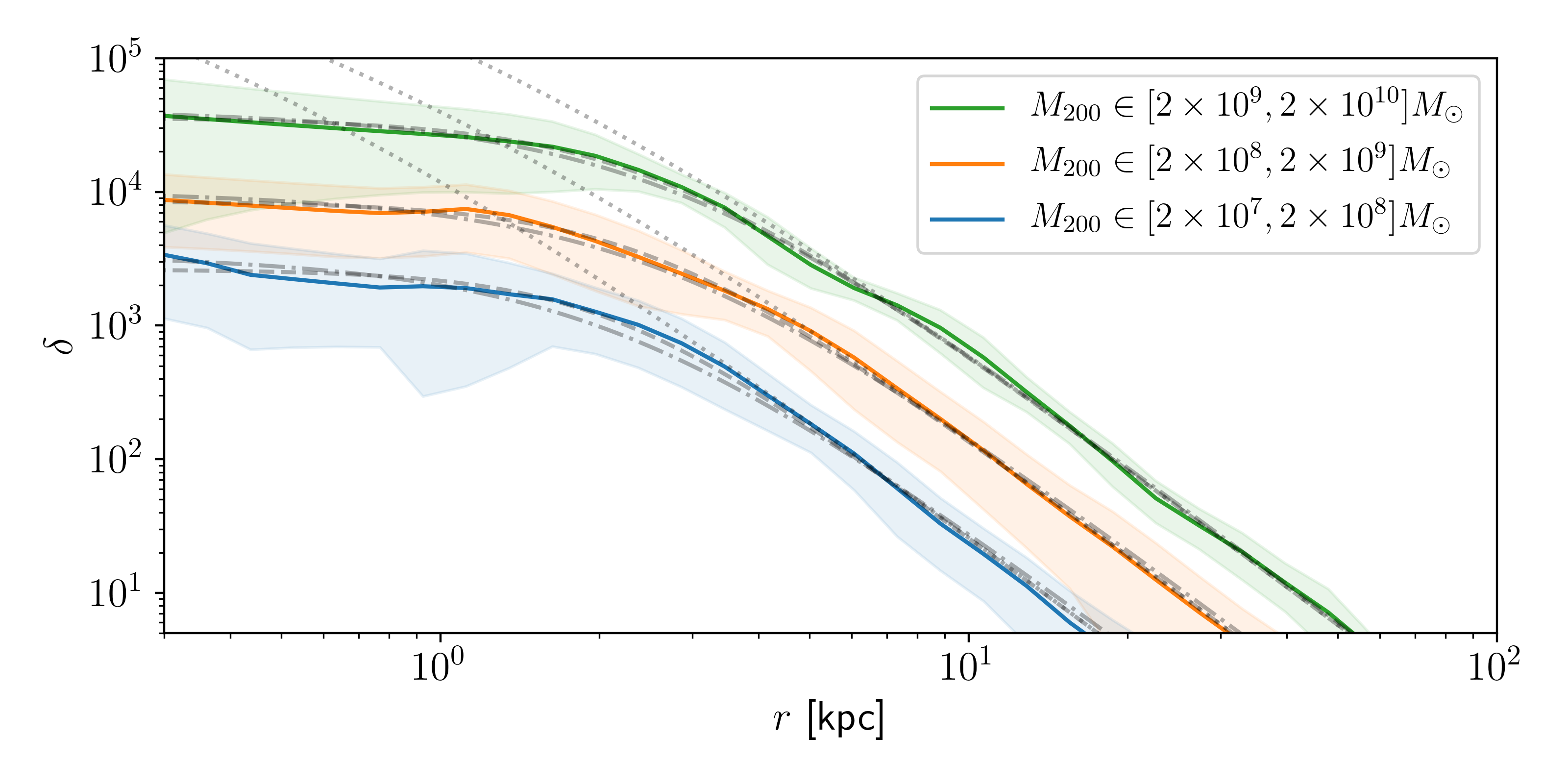}}

    \caption{The mean (\textit{solid}) and standard deviation (\textit{shaded}) of binned SIBEC-DM halo profiles with $R_c=1\text{kpc}$ at $z=0.5$. The binned profiles are fitted to NFWc (\textit{dashed}) and Burkert (\textit{dash-dotted}). The fits of the NFW profile to the halo envelopes are also shown (\textit{dotted}).}
\label{fig:NFWc_Burkert_profile_fits}
\end{figure}

Both NFWc and Burkert fitted to binned SIBEC-DM halos with $R_c=1\text{kpc}$ at $z=0.5$ are shown in figure~\ref{fig:NFWc_Burkert_profile_fits}.
The two profiles yield slightly different results for the core radius. Burkert
is a bit steeper than NFWc near the core, and therefore generally gives a core radius $r_c$ of around half compared to NFWc (according to our definition $\delta(r_c) = 0.5\delta(0)$). The core density $\delta_c$ is also about twice as high for Burkert compared to NFWc. In the following we use the Burkert profile, even though the fit using NFWc is better, for two reasons; it is a simpler function, and we can readily compare the fitting parameters of the SIBEC-DM halos to observations, because the Burkert profile has previously been used for the DM component of nearby galaxies in the SPARC dataset \citep{Li2020} and the dwarf spheroidal (dSph) satellites of the Milky Way \citep{Salucci2012}.

In order to investigate the general trend in the properties of the SIBEC-DM halos in our simulations, and compare these to observations, we introduce scaling functions for the core radius $r_c$, the central density $\delta_c$, and the mass enclosed inside the core radius, $M_c$;
\begin{equation}
\label{eq:rc_fitting_func}
    r_c(M_{200}) = r_{c,10}\Bigg(\frac{M_{200}}{10^{10} M_{\odot}}\Bigg)^{\alpha},
\end{equation}
\begin{equation}
\label{eq:deltac_fitting_func}
    \delta_c(M_{200}) = \delta_{c,10}\Bigg(\frac{M_{200}}{10^{10} M_{\odot}}\Bigg)^{\beta},
\end{equation}

\begin{equation}
\label{eq:Mc_fitting_func}
    M_c(M_{200}) = M_{c,10}\Bigg(\frac{M_{200}}{10^{10} M_{\odot}}\Bigg)^{\gamma}.
\end{equation}
These are fitted using Theil-Sen regression \citep{Theil1950,Sen1968}. This method finds the slope $m$ and the $y$-intercept $b$ of the linear function $y(x) = b+mx$ (which eqs. \eqref{eq:rc_fitting_func}, \eqref{eq:deltac_fitting_func}, and \eqref{eq:Mc_fitting_func} are in log-space) by first computing the median of slopes between all pairs of points $(y_i, x_i)$, which gives $m$, and then finding the median of $y_i-mx_i$ for all points, which gives $b$. This fitting procedure is robust against outliers, and provides a simple measure of the variation in $m$ and $b$ present in the data through e.g. the 1st and 3rd quartiles of the pairwise slopes and individual $y$-intercepts. Tables listing the scaling parameters fitted using this procedure are included in appendix~\ref{app:fitted_scaling_relation_parameters} for several redshifts.

\begin{figure*}
    \centering
    \subfigure{\includegraphics[width=0.33\linewidth]{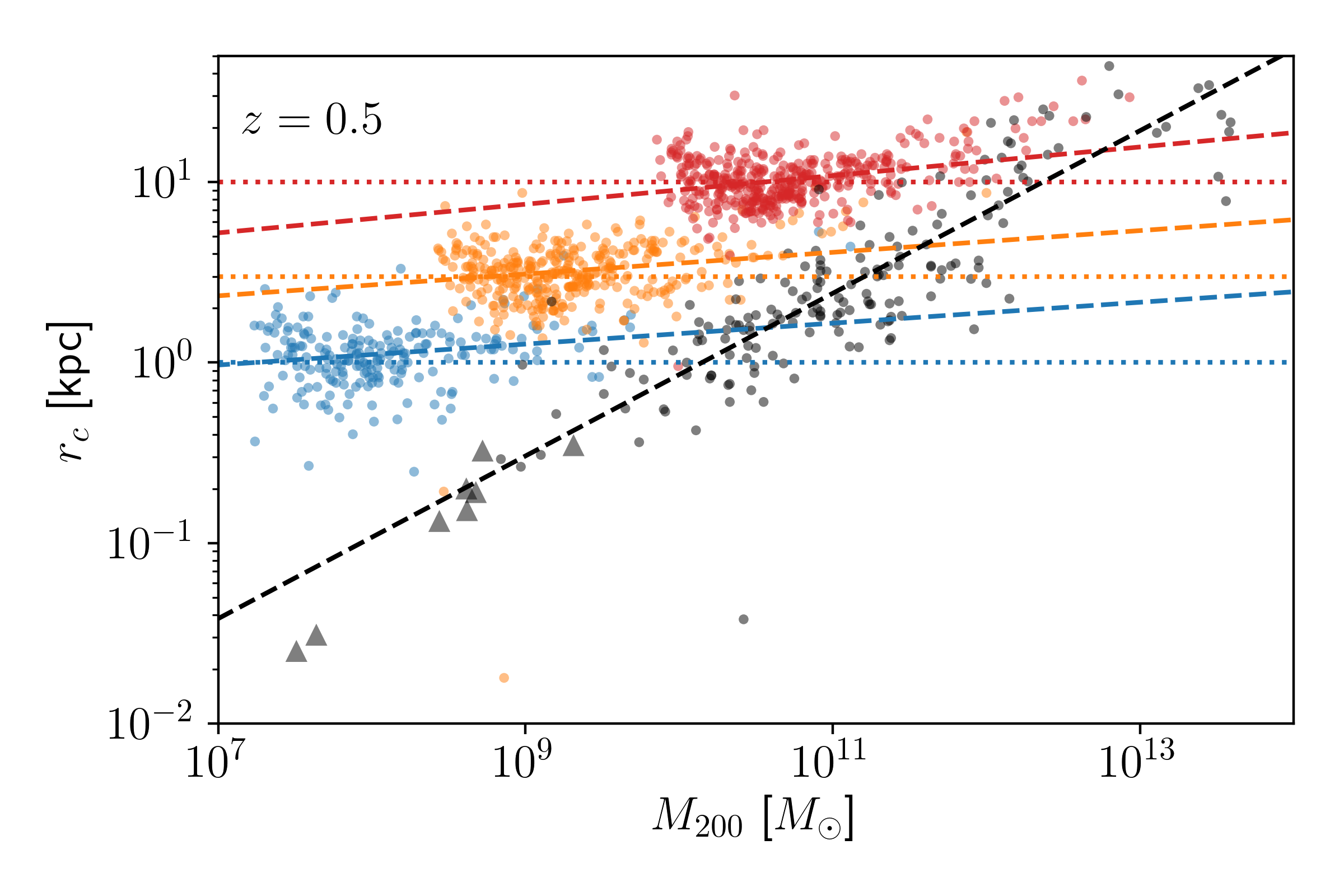}}
    \subfigure{\includegraphics[width=0.33\linewidth]{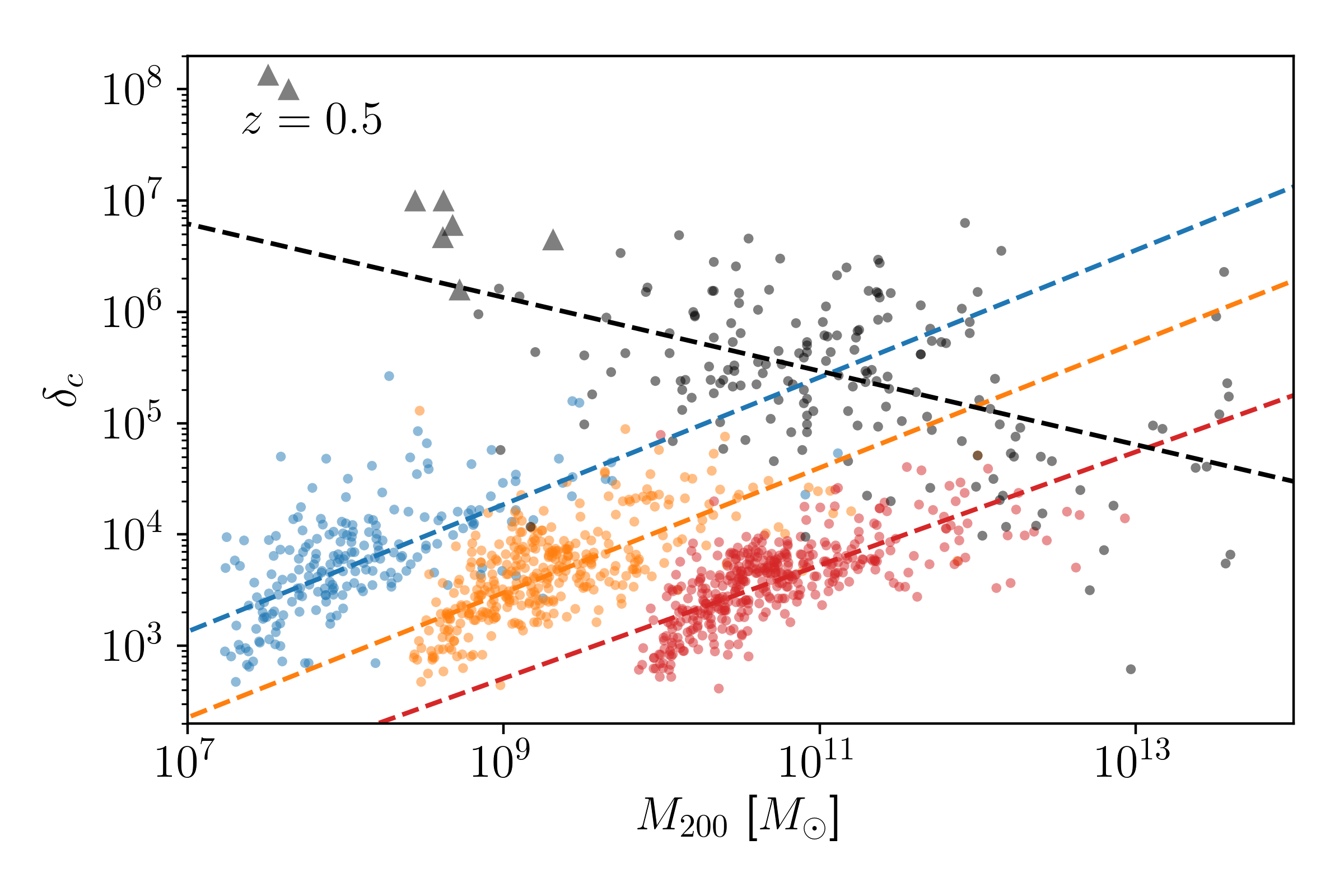}}
    \subfigure{\includegraphics[width=0.33\linewidth]{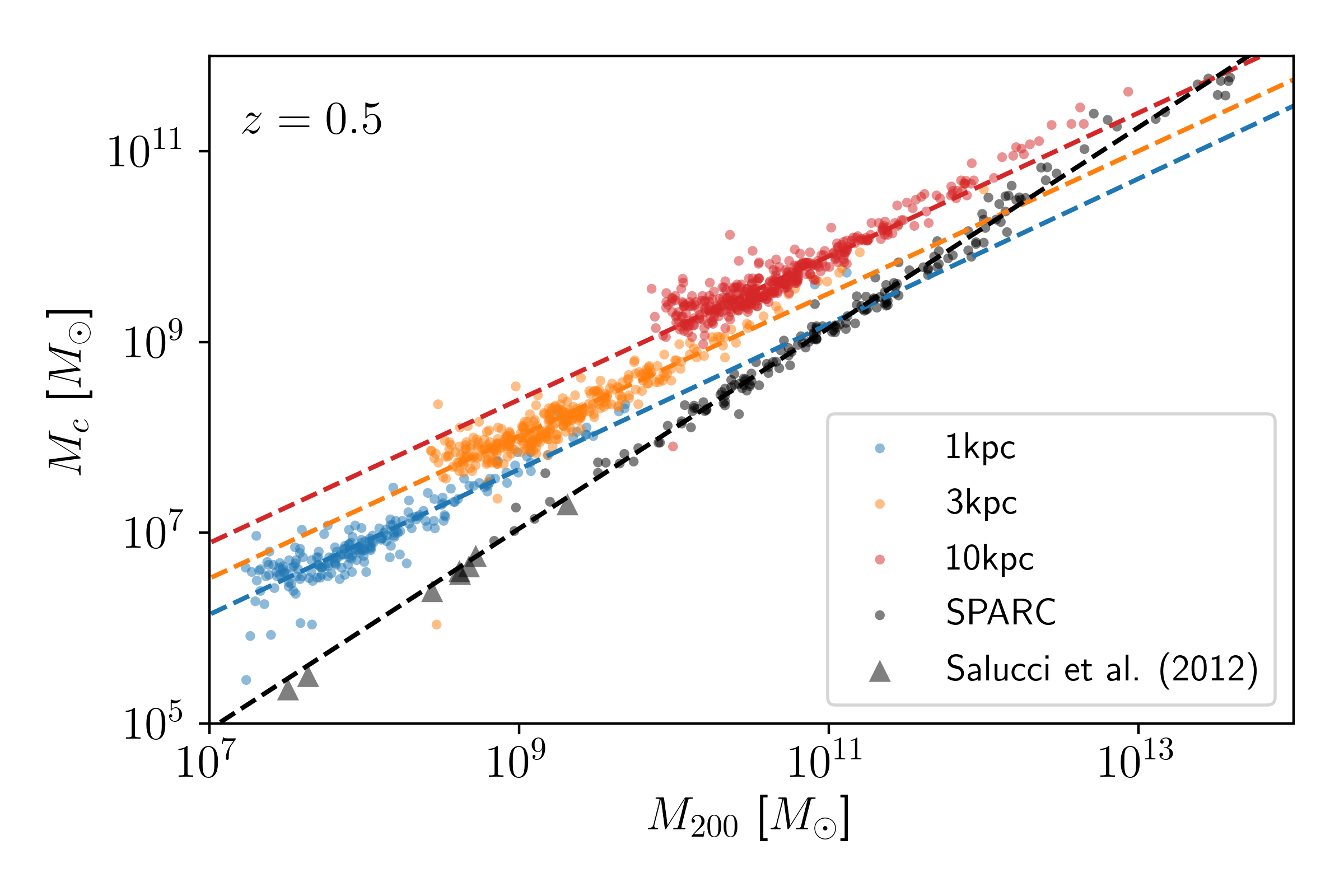}}
    \subfigure{\includegraphics[width=0.33\linewidth]{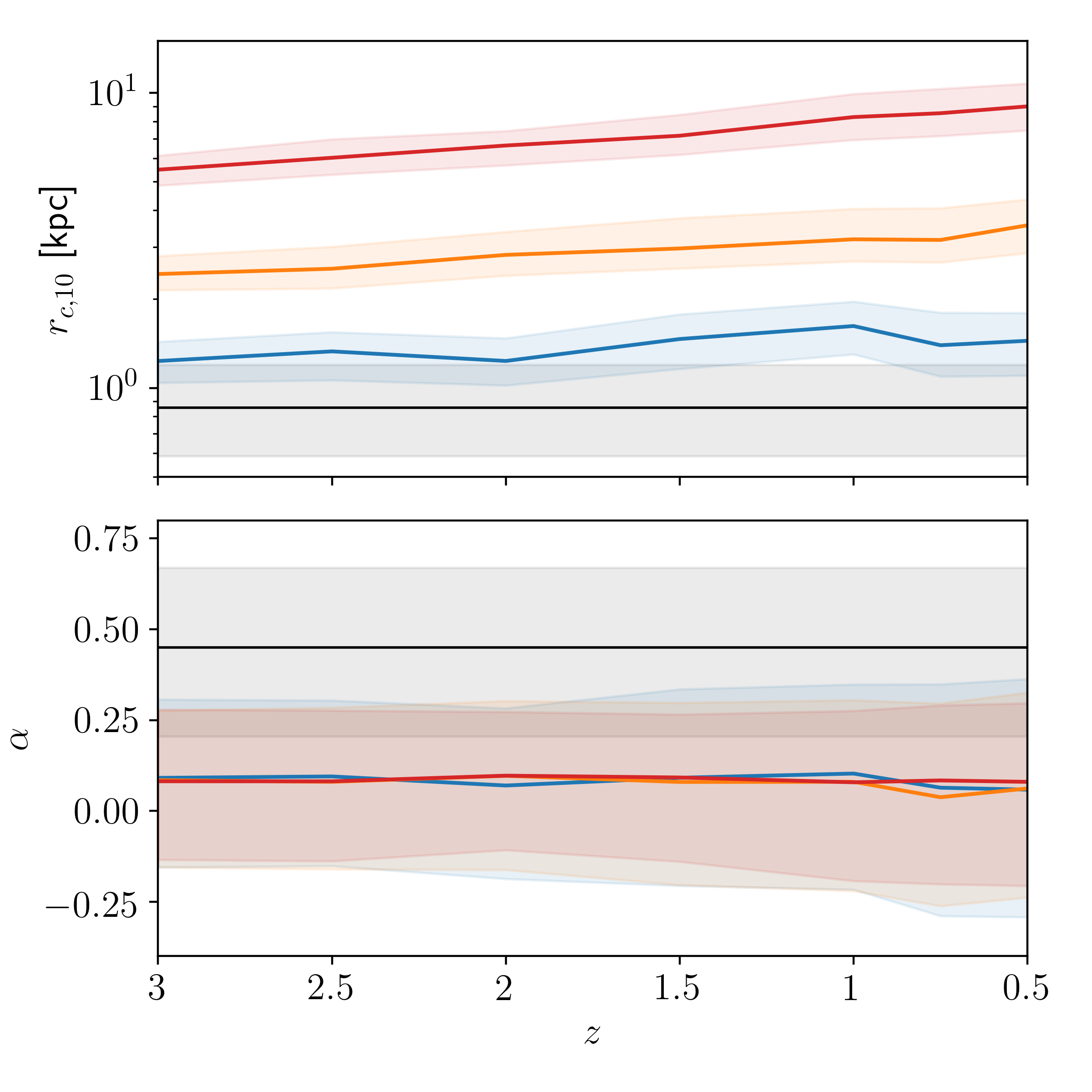}}
    \subfigure{\includegraphics[width=0.33\linewidth]{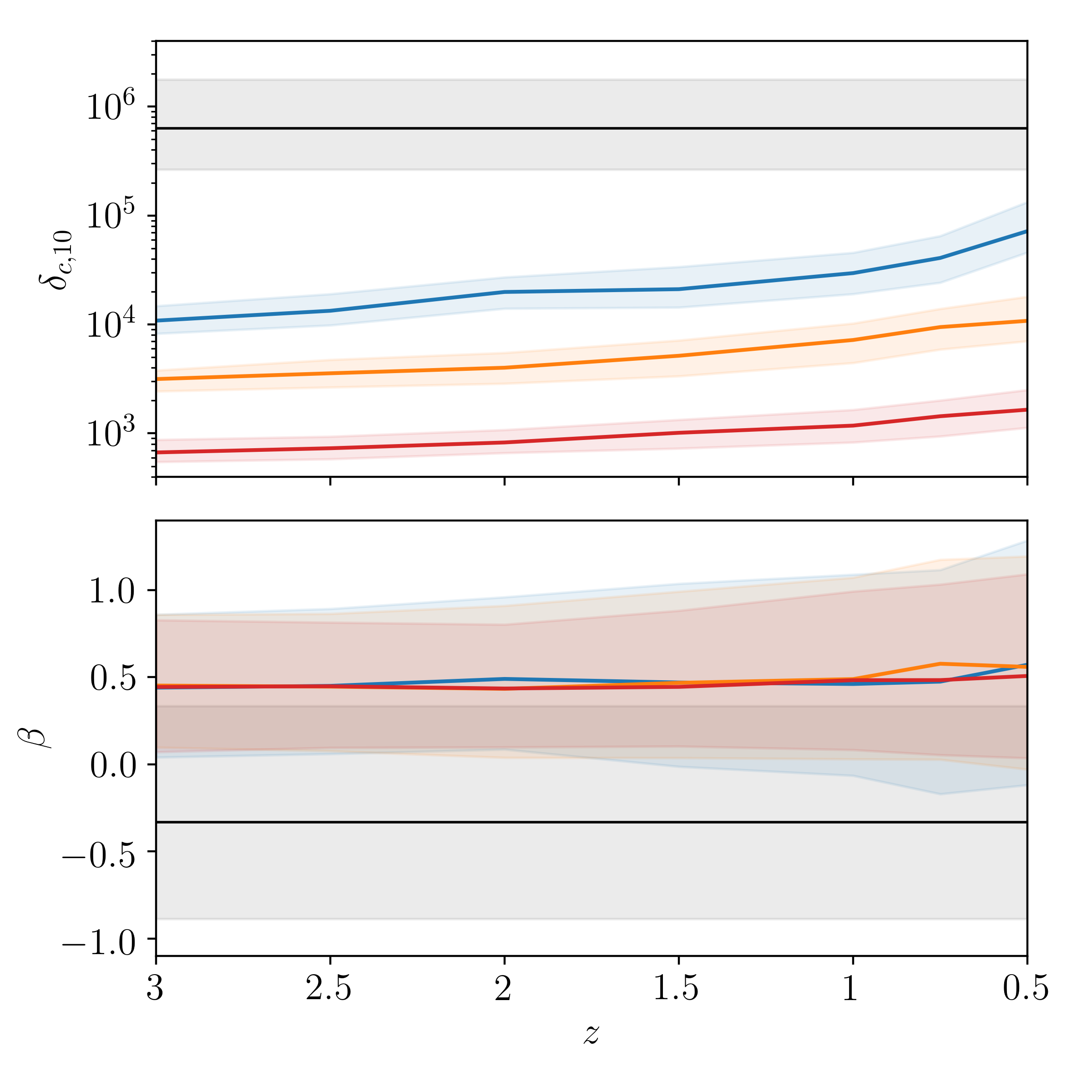}}
    \subfigure{\includegraphics[width=0.33\linewidth]{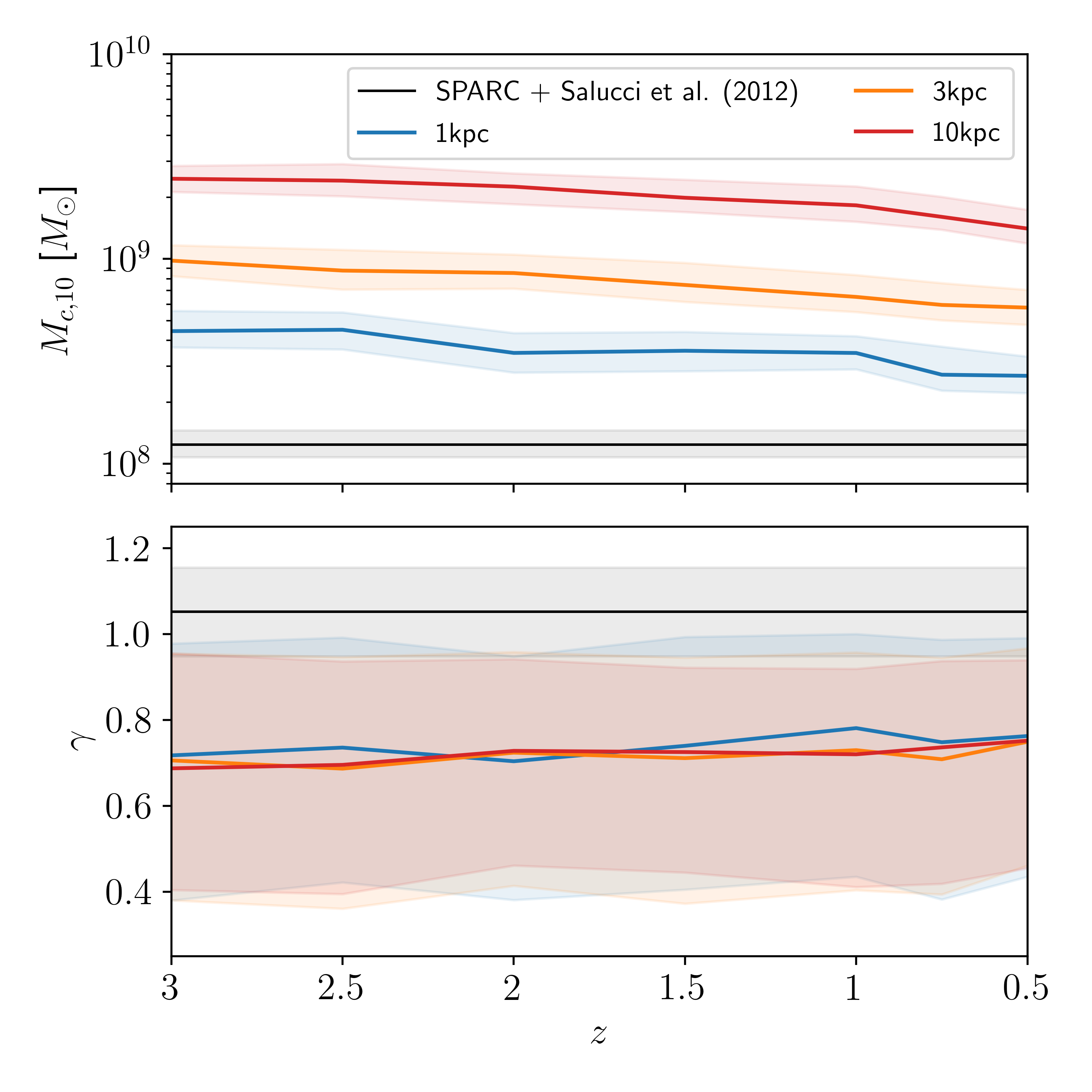}}

    \caption{The core radii $r_c$ (\textit{left}), core densities $\delta_c$ (\textit{middle}), and core masses $M_c$ (\textit{right}) versus $M_{200}$ for halos in cosmological simulations of SIBEC-DM with $R_c=1\text{kpc}$, $3\text{kpc}$, and $10\text{kpc}$, using the Burkert profile. The upper plots show the scatter of halos at $z=0.5$, with the fitted scaling eqs. \eqref{eq:rc_fitting_func}, \eqref{eq:deltac_fitting_func}, and \eqref{eq:Mc_fitting_func} shown in dashed lines, compared to the SPARC dataset \citep{Li2020} and Milky Way dSphs \citep{Salucci2012}. The lower plots show the fitted median (\textit{solid}) and the 1st and 3rd quartiles (\textit{shaded}) at several redshifts using Theil-Sen regression. The results from SPARC and the dSphs are also shown. In the scatter plot for $r_c$ the core radii $R_c$ are indicated in dotted colored lines.}
\label{fig:simulation_fitted_parameters}
\end{figure*}

\begin{figure}
    \centering
    \includegraphics[width=0.98\linewidth]{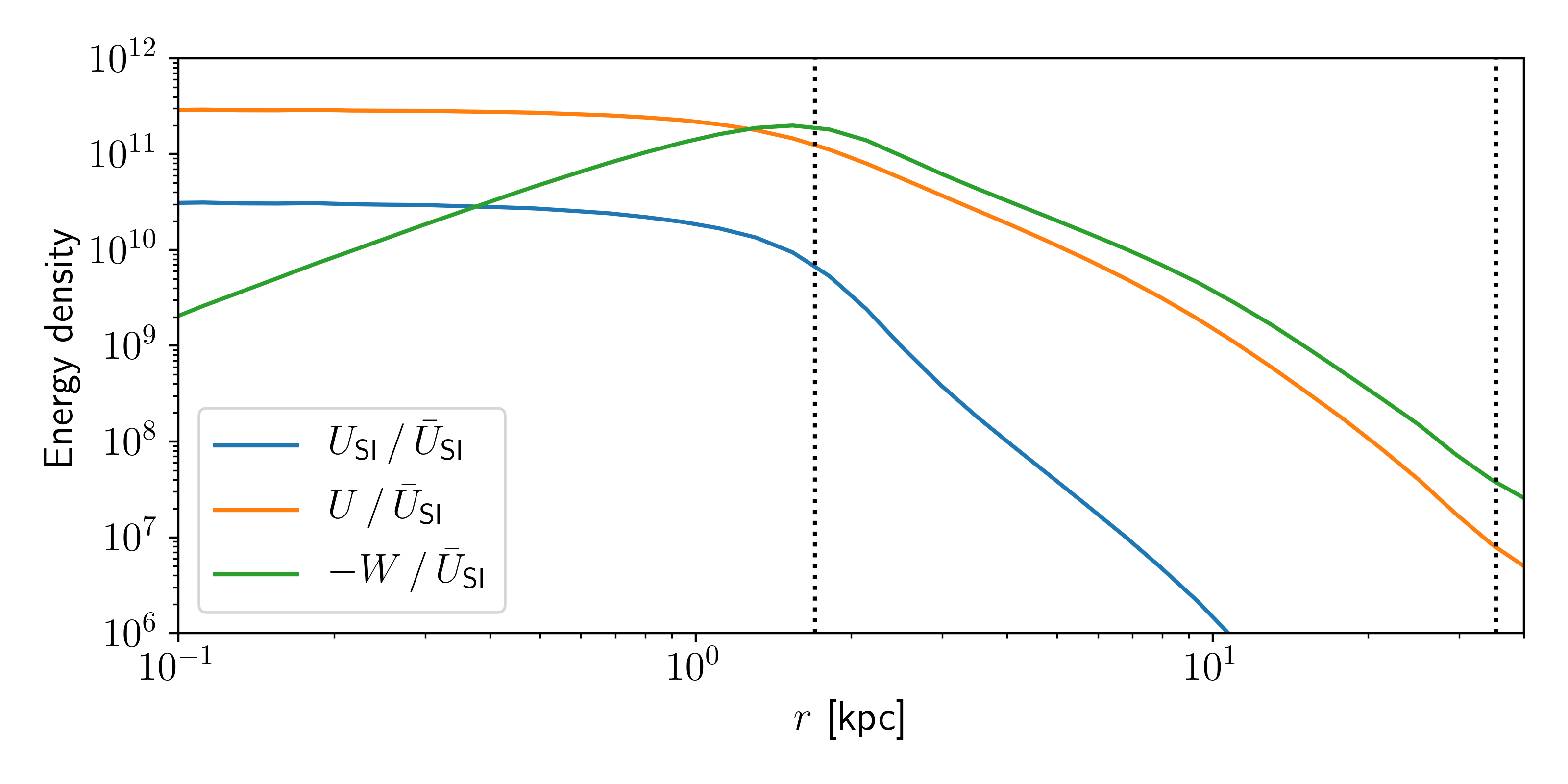}
    \includegraphics[width=0.98\linewidth]{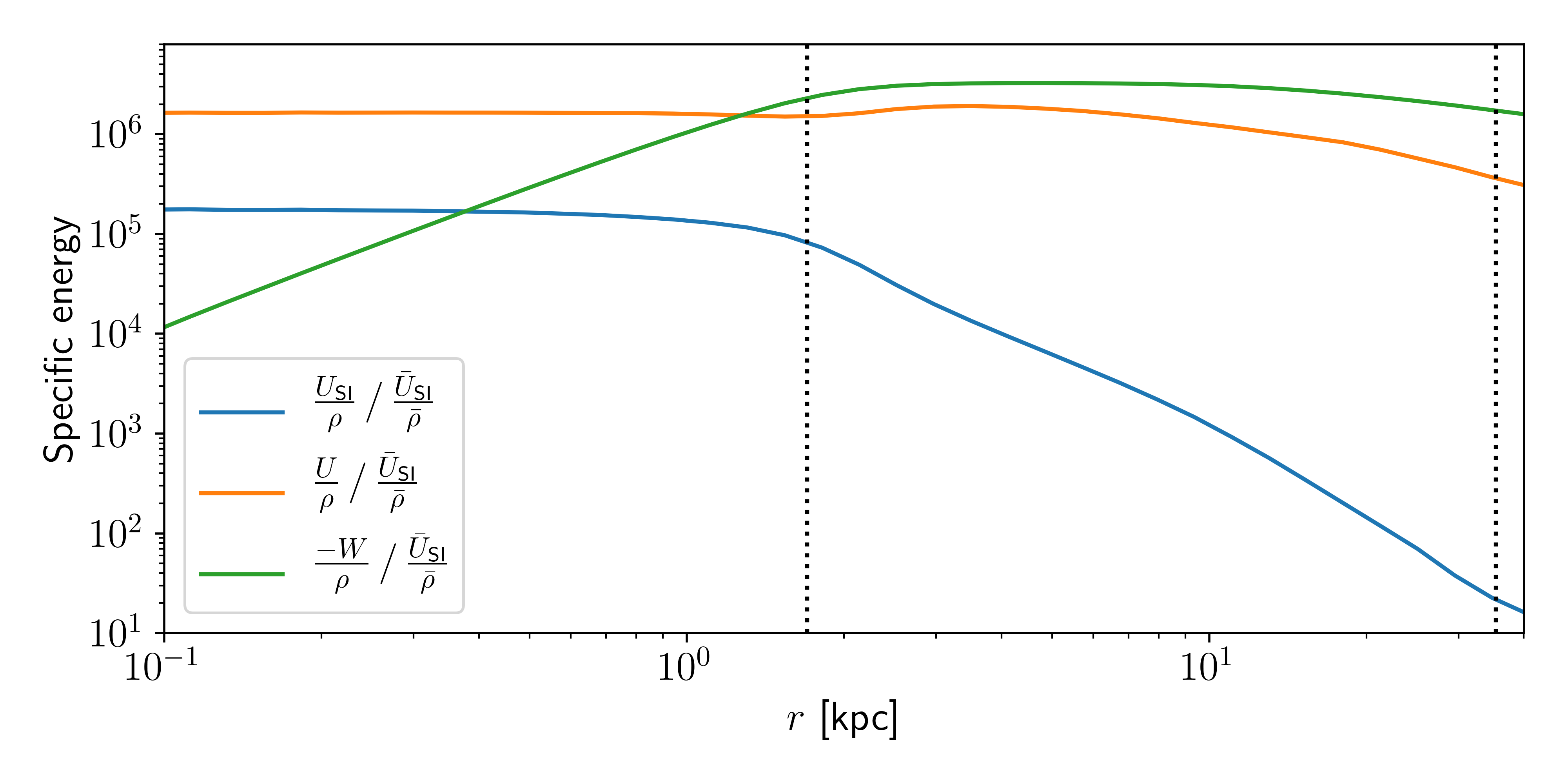}
    \includegraphics[width=0.98\linewidth]{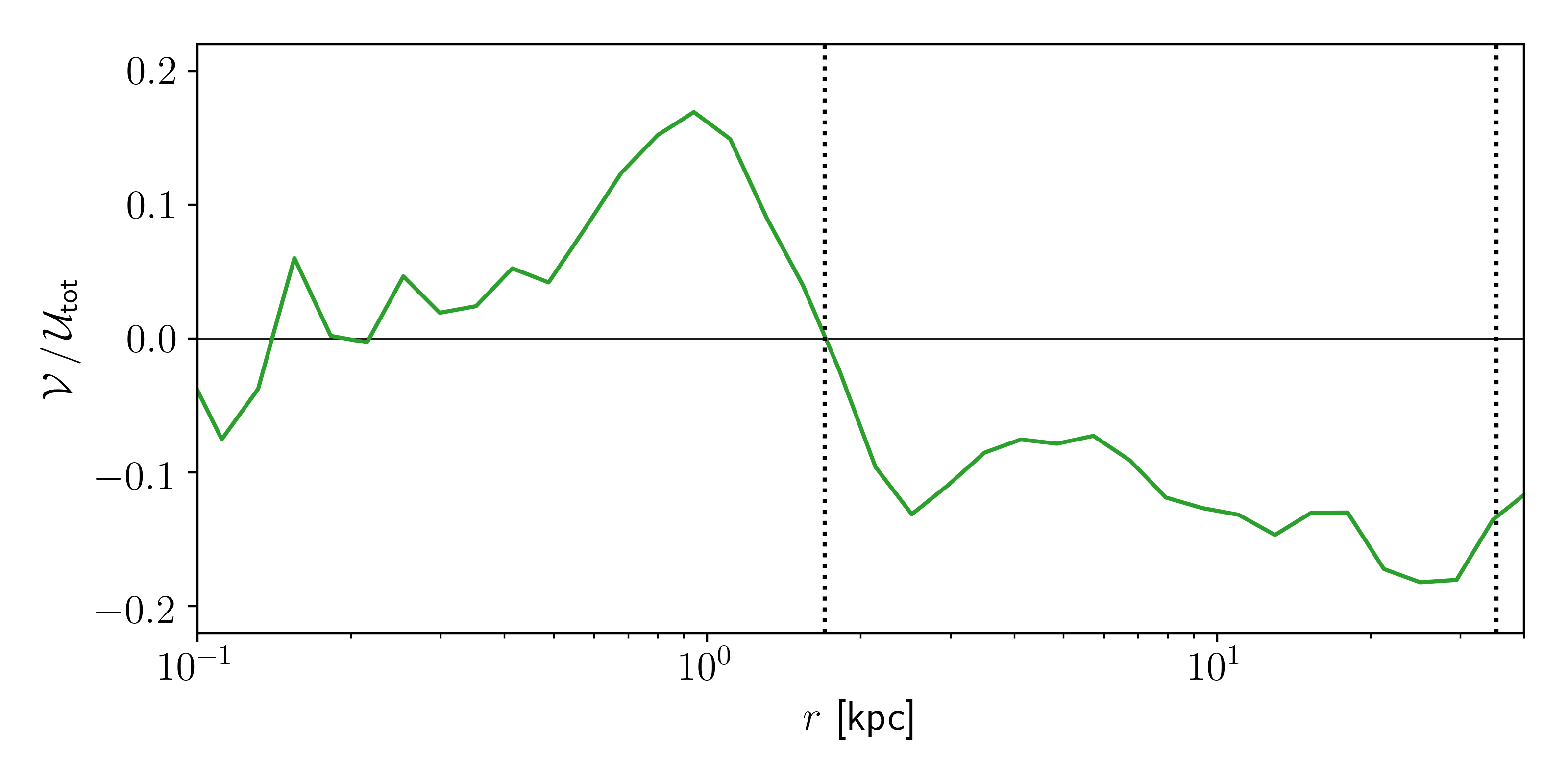}

    \caption{The top panel shows the self-interaction energy density $U_{\text{SI}}$, thermal energy density $U$, and gravitational potential energy density $W$ of a sample halo with $R_c=1\text{kpc}$ and $M_{200}=3\times 10^{9}M_{\odot}$, relative to the interaction energy at the background level $\bar{U}_{\text{SI}}$. The middle panel shows instead the specific energy, and the bottom panel the cumulative virial. The core radius $r_c$ and halo radius $r_{200}$ are indicated with the inner and outer dotted vertical lines, respectively.}
\label{fig:halo_energy_contributions}
\end{figure}

The properties of the Burkert profile fitted to SIBEC-DM halos are shown in figure~\ref{fig:simulation_fitted_parameters} for $R_c = 1\text{kpc}$, $3\text{kpc}$, and $10\text{kpc}$, along with the above scaling relations. The fit to the nearby galaxy rotation curves in the SPARC dataset \citep{Li2020} and Milky Way dSphs \citep{Salucci2012} are also included in figure~\ref{fig:simulation_fitted_parameters}.
While the halo catalogues in the simulation snapshots only span 2-3 orders in magnitude in mass and number 200-400 halos due to the limited volume of the simulated box, and the fitted parameters of the scaling functions show some time-dependence (the prefactors more so than the exponents), there are a number of interesting features to note. First of all, the halo core radii $r_c$ are scattered near $R_c$, and the dependence of $r_c$ on $M_{200}$ is nearly constant, as expected from hydrostatic considerations. Specifically, it is found that $r_c\sim M_{200}^{\alpha}$ with $\alpha\approx 0.05-0.1$, with the 1st and 3rd quartiles of $\alpha$ also consistent with $\alpha=0$. The SIBEC-DM cores are generally larger compared to what we find from eq. \eqref{eq:SIBEC_DM_hydrostatic_profile}, a consequence of the additional pressure and that the halo envelopes are rotationally supported, following the NFW profile. In fact, the thermal-like energy due to the hidden dynamics on the de Broglie-scale dominates the central regions of the SIBEC-DM halos. This can be seen in figure~\ref{fig:halo_energy_contributions}, where the self-interaction, thermal, and gravitational potential energy densities of a sample halo are shown, with
\begin{equation}
    W = -\frac{GM(<r)\rho}{r}.
\end{equation}
As structure forms, the SIBEC-DM fluid is heated as it is accreted onto the growing halos, the kinetic energy of the in-falling matter being converted into effective thermal energy. The result are halos with a thermal energy much larger than the interaction energy throughout the halo, even in the core, where it is around a factor 10 times larger and dominates the overall dynamics. Additionally, the thermal energy is nearly isothermal. At the edge of the core the potential energy becomes larger than the internal energies, and the halo transitions to the NFW-like envelope.

We illustrate with the sample halo in figure~\ref{fig:halo_energy_contributions}, which is relatively isolated in the simulated box, that the SIBEC-DM halos in the simulations achieve approximate virial equilibrium. The virial $\mathcal{V}$ satisfies, under the assumption of spherical symmetry \citep{Dawoodbhoy2021},
\begin{equation}
    \mathcal{V} = 2\mathcal{T} + \mathcal{W} + 2\mathcal{U} + 3\mathcal{U}_{\text{SI}} + \mathcal{S} + \mathcal{S}_{\text{SI}} = 0,
\end{equation}
where the various terms are cumulative energy contributions and surface terms,
\begin{equation}
    \mathcal{T} = \frac{1}{2}\int v^2\text{d}M - 2\pi r^2\Bigg(\rho r v^2 + \frac{1}{2}r^2\frac{\partial(\rho v)}{\partial t} \Bigg),
\end{equation}
\begin{equation}
    \mathcal{W} = -\int \frac{GM}{r}\text{d}M,
\end{equation}
\begin{equation}
    \mathcal{U} = \int \frac{U}{\rho}\text{d}M, \quad \quad \mathcal{U}_{\text{SI}} = \int \frac{U_{\text{SI}}}{\rho}\text{d}M,
\end{equation}
\begin{equation}
    \mathcal{S} = -4\pi r^3 P,\quad\quad \mathcal{S}_{\text{SI}} = -4\pi r^3 P_{\text{SI}},
\end{equation}
and $\mathcal{U}_{\text{tot}} = \mathcal{U} + \mathcal{U}_{\text{SI}}$.

\begin{figure}
    \centering
    \includegraphics[width=0.98\linewidth]{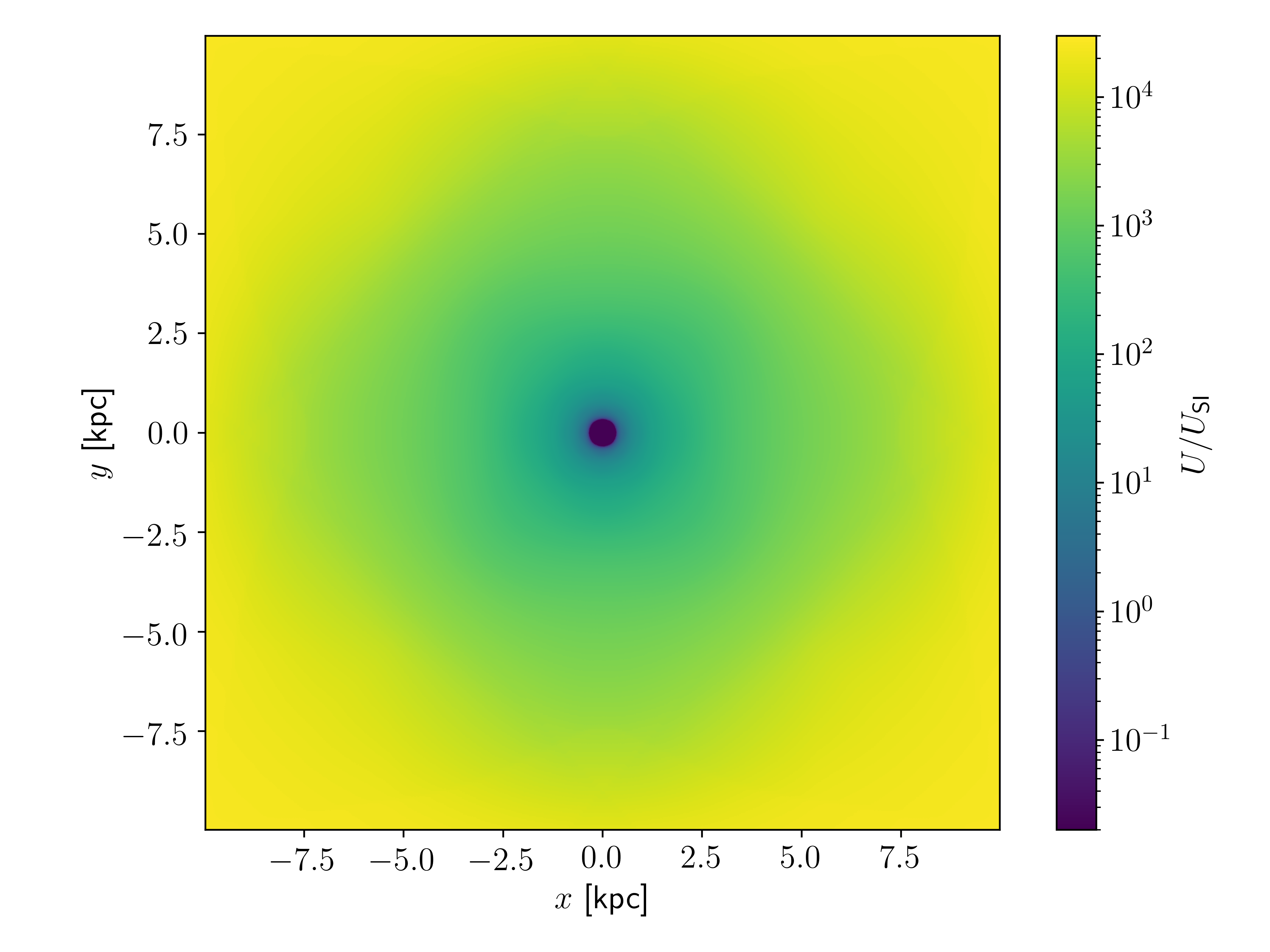}
    \includegraphics[width=0.98\linewidth]{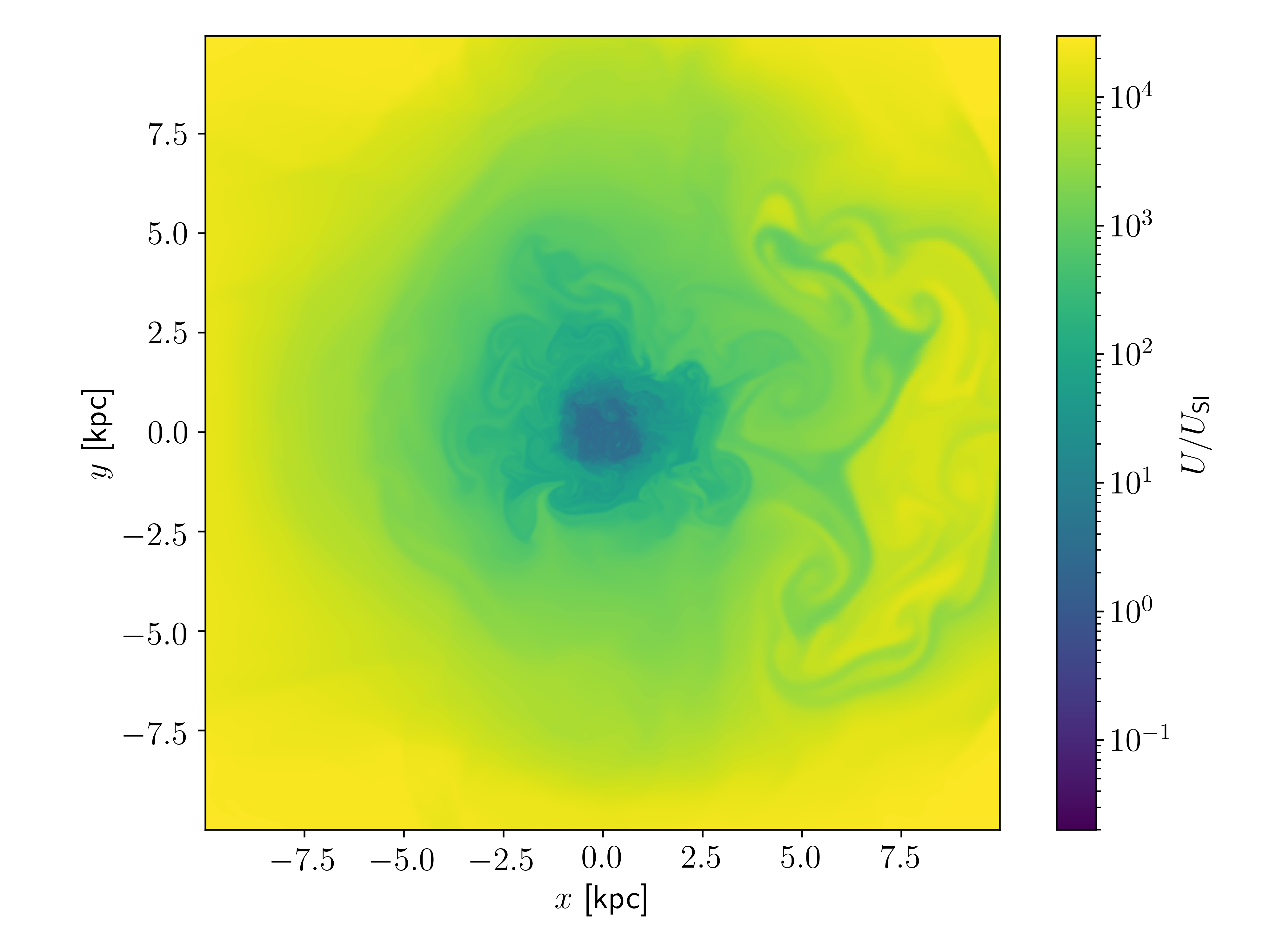}
    
    \caption{Final distributions of $U/U_{\text{SI}}$ in 2D simulations with symmetrical (\textit{upper}) and asymmetrical (\textit{lower}) collapse, centered on the density peak. In the asymmetrical case, the second smaller over-density was initially located to the right. The minimum in the symmetrical case is around $U/U_{\text{SI}} \approx 2\times 10^{-2}$, while in the asymmetrical case its is $U/U_{\text{SI}} \approx 2$.}
\label{fig:2d_sims_U_over_USI}
\end{figure}

The domination of thermal energy in the halo cores is in contrast to the spherically symmetric 1D simulations of \citet{Dawoodbhoy2021}, who instead found the thermal energy to fall below the interaction energy inside the core. In their simulations many of the same processes takes place, but the cores are not heated and are dominated by the self-interaction pressure. Instead our 3D simulations look more like the isothermal profiles of \citet{Dawoodbhoy2021} with the thermal Jeans' length on the order of $R_c$. We attribute the difference between the results of our simulations to the absence of mixing in spherical symmetry. In the 1D simulations the halos form by the accretion of spherically symmetric shells that collapse onto the halos and are heated as they are slowed down, resulting in an accretion shock expanding outwards as shells collide, but do not cross. The cores themselves experience little heating as the fluid is decelerated smoothly and without a shock due to the self-interaction pressure. In fully 3D simulations the halos form as clumps --- not perfect shells --- of matter merge, which perturb the halos and causes the outer shock heated layers to mix with the interior. We demonstrate this difference between symmetrical and asymmetrical collapse using 2D simulations; one in which an initial stationary over-density centered in the simulated box collapses symmetrically; and another with a second smaller over-density offset at $\bm{x}_2$ that merges with the first in an asymmetrical manner. The initial density field in the two scenarios are $\rho(\bm{x}) = \rho_0 + \rho_0\Delta_1 e^{-(|\bm{
x}|/R_1)^2}$, and $\rho(\bm{x}) = \rho_0 + \rho_0\Delta_1 e^{-(|\bm{
x}|/R_1)^2} + \rho_0\Delta_2 e^{-(|\bm{
x}-\bm{x_2}|/R_2)^2}$, and we use a periodic box of size $L=40\text{kpc}$ with $R_c=1\text{kpc}$, $R_1 = 5\text{kpc}$, $R_2 = 2\text{kpc}$, $\Delta_1 = \Delta_2 = 99$, $|\bm{x}_2| = 10\text{kpc}$, and $\rho_0 = \Omega_{m0}\rho_{c0}$, with the initial ratio $\zeta=P/P_{\text{SI}}=10^{-1}$. The simulations were run for $20 t_{\text{dyn}}$, where $t_{\text{dyn}} = 1/\sqrt{2\pi G\rho_0 \Delta_1}$ is the dynamical timescale of the central over-density. The final distributions of $U/U_{\text{SI}}$ are shown in figure~\ref{fig:2d_sims_U_over_USI}. In the symmetrical case, the thermal energy is dominant everywhere except in the core, whereas in the asymmetrical case the merging with the second over-density results in a strong mixing of the fluid layers, causing shock-heated fluid in the region outside the core to penetrate into the core.

\begin{figure*}
    \centering
    \includegraphics[width=0.495\linewidth]{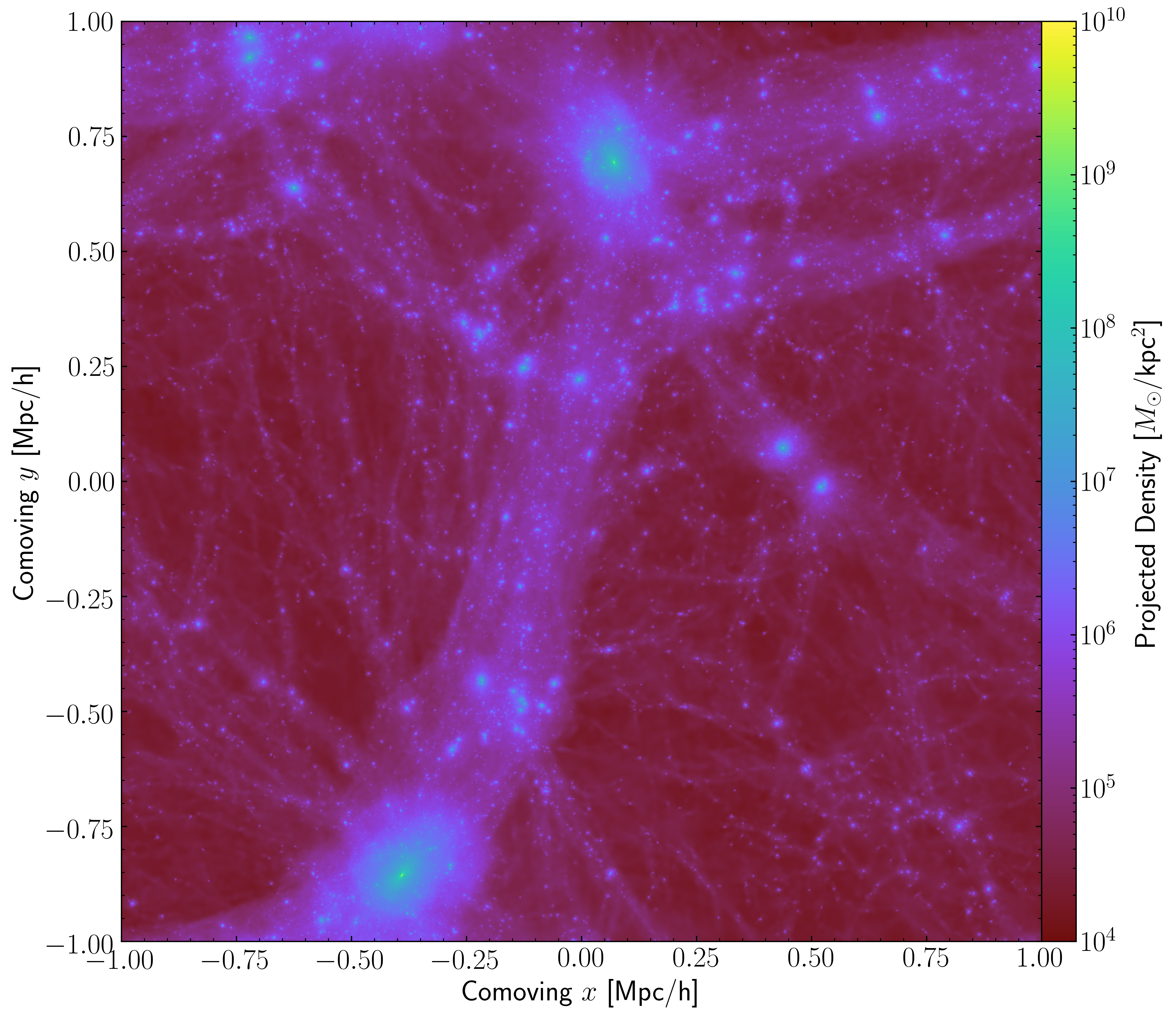}
    \includegraphics[width=0.495\linewidth]{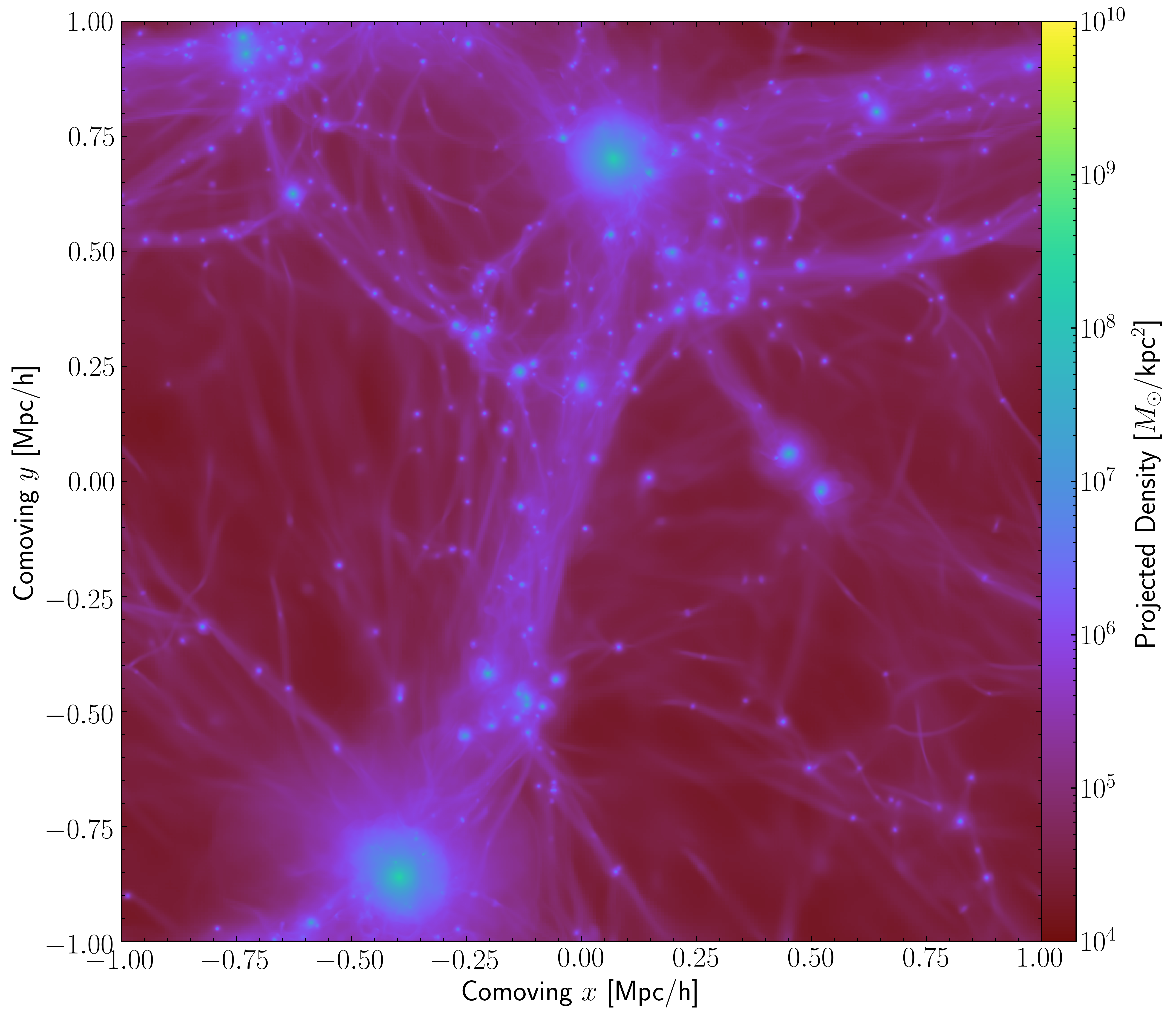}
    \includegraphics[width=0.495\linewidth]{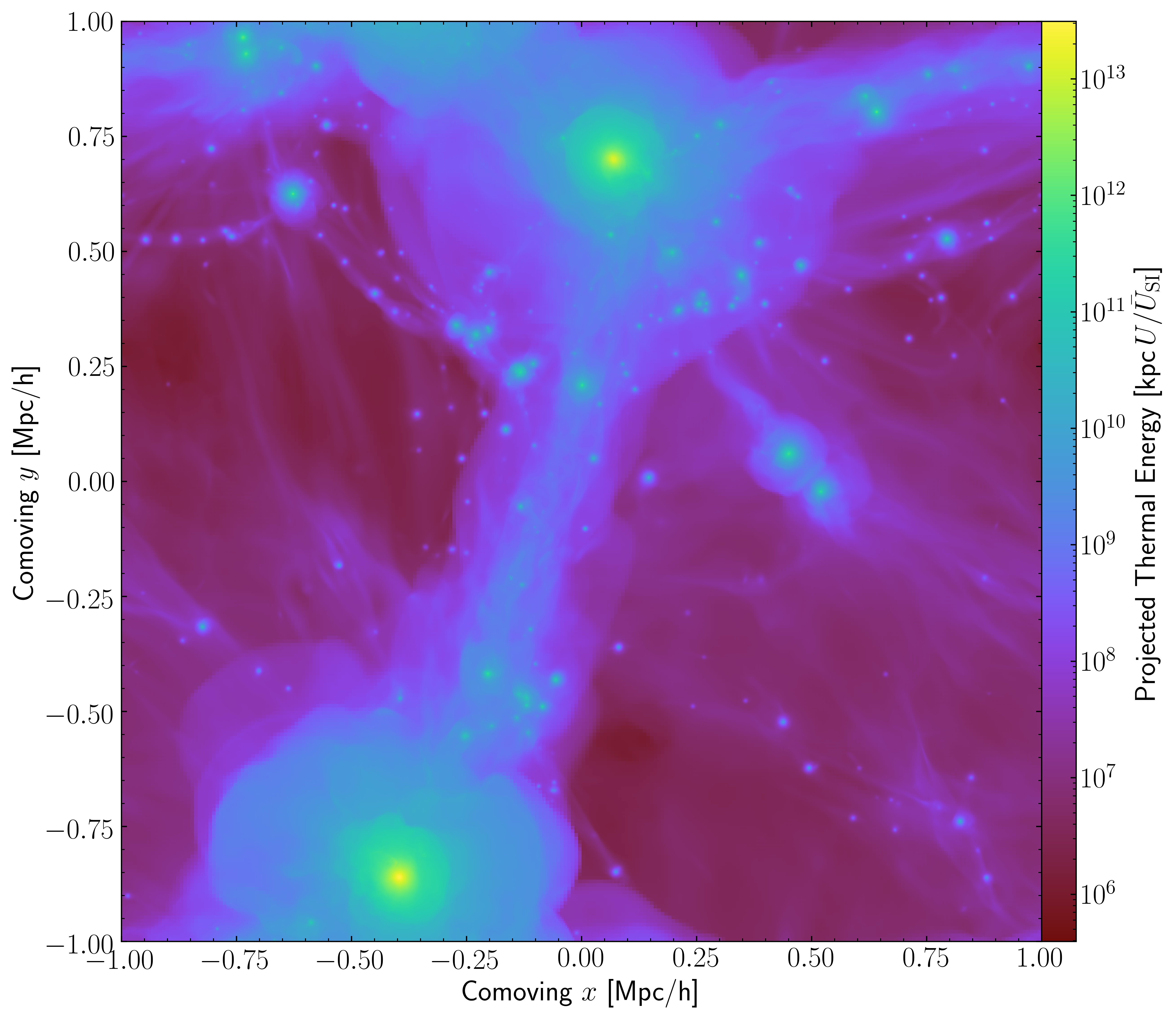}
    \includegraphics[width=0.495\linewidth]{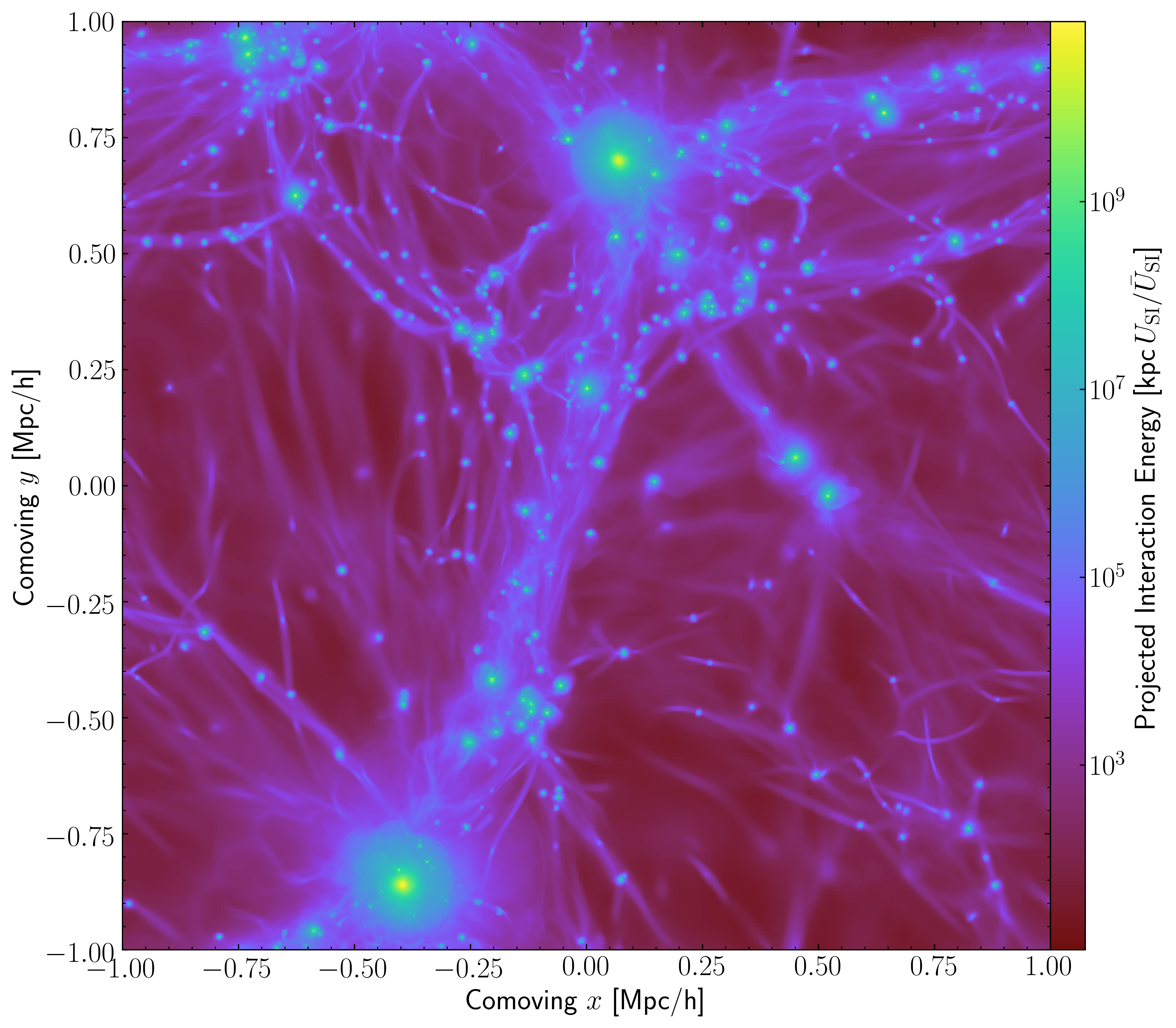}

    \caption{Projection plots of the DM density in cosmological simulations with $L=2\text{Mpc}/h$ at $z=0.5$ SIBEC-DM with $R_c=1\text{kpc}$ (\textit{upper right}) and CDM with the same initial conditions, but without a cut-off (\textit{upper left}). The SIBCE-DM internal energies for the same snapshot showing the effective thermal energy $U$ (\textit{lower left}) and the self-interaction energy $U_{\text{SI}}$ (\textit{lower right}).}
\label{fig:simulation_projections_plots}
\end{figure*}

\begin{figure}
    \centering
    \includegraphics[width=0.98\linewidth]{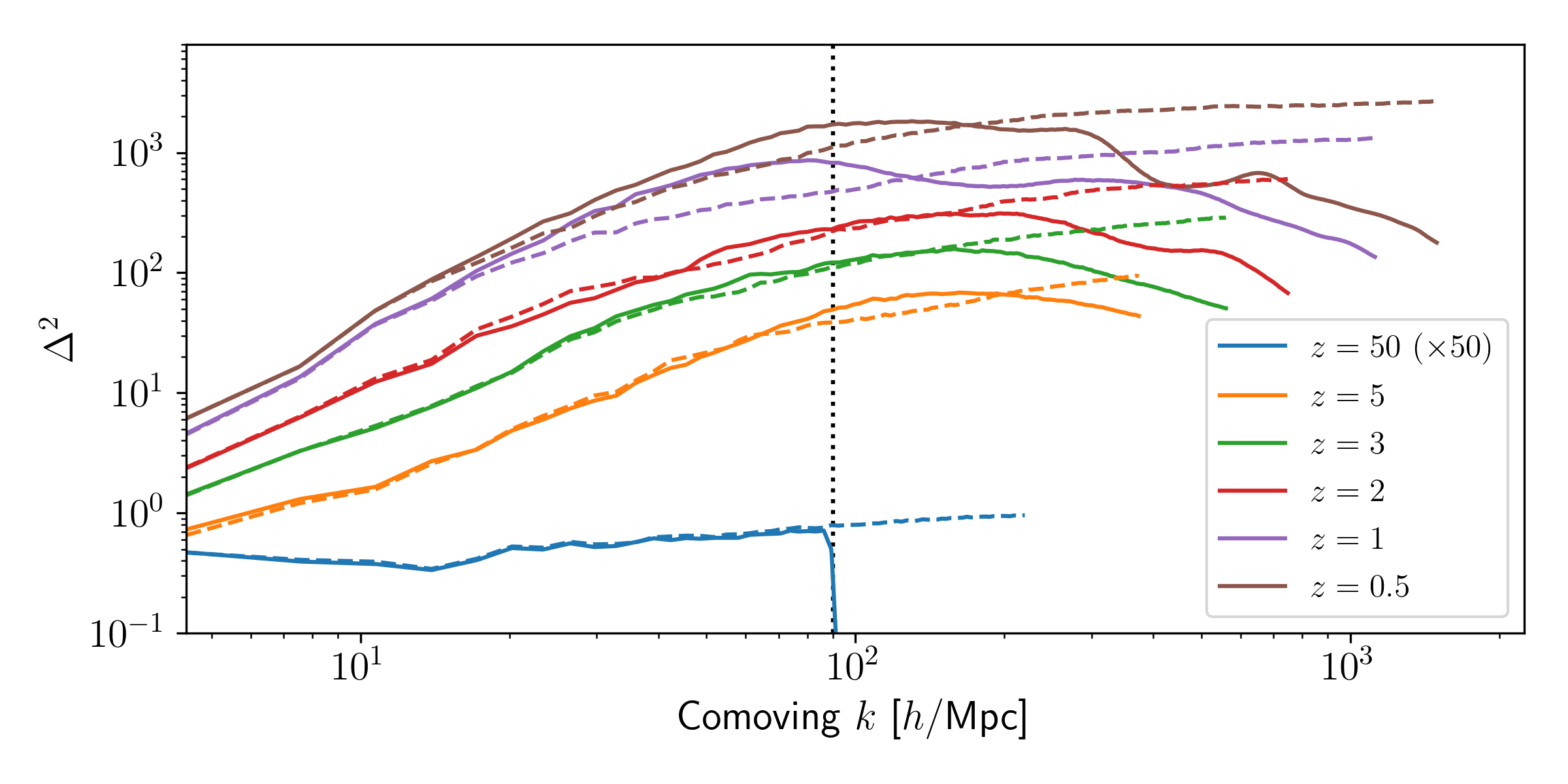}

    \caption{The matter power spectrum for cosmological simulations $L=2\text{Mpc}/h$ at $z=0.5$ for SIBEC-DM with $R=1\text{kpc}$ (\textit{solid}) and CDM (\textit{dashed}). The comoving cut-off $k_{\text{cut}}$ is indicated with a dotted vertical line, and the spectrum at $z=50$ is multiplied by $50$ in the figure.}
\label{fig:1kpc_CDM_matter_power_spectra}
\end{figure}

In figure~\ref{fig:simulation_projections_plots} the internal energies of SIBEC-DM with $R_c=1\text{kpc}$ at $z=0.5$ are shown, where the extended envelopes of high thermal energies compared to the interaction energy are clearly seen around the collapsed structures. In the same figure the density field is compared to standard CDM, which illustrates the smoothing of the DM density field, especially in the halo cores. The corresponding matter power spectrum is shown in figure~\ref{fig:1kpc_CDM_matter_power_spectra} at several redshifts, as well as CDM with the same initial conditions, but without the cut-off, with
\begin{equation}
    \Delta^2(k) = \frac{k^3P(k)}{2\pi^2},
\end{equation}
where
\begin{equation}
    (2\pi)^3 P(k)\delta^3(\bm{k}-\bm{k}') = \braket{\hat{\delta}(\bm{k})\hat{\delta}(\bm{k}')},
\end{equation}
and
\begin{equation}
    \hat{\delta}(\bm{k}) = \int \text{d}^3x \,\delta(\bm{x})e^{-i\bm{k}\cdot\bm{x}},
\end{equation}
with $\delta^3$ being the 3D Dirac delta function. At the largest scales the SIBEC-DM modes grow like CDM, while at smaller scales SIBEC-DM is suppressed compared to CDM due to the presence of fluid pressure. 

\begin{figure}
    \centering
    \includegraphics[width=0.98\linewidth]{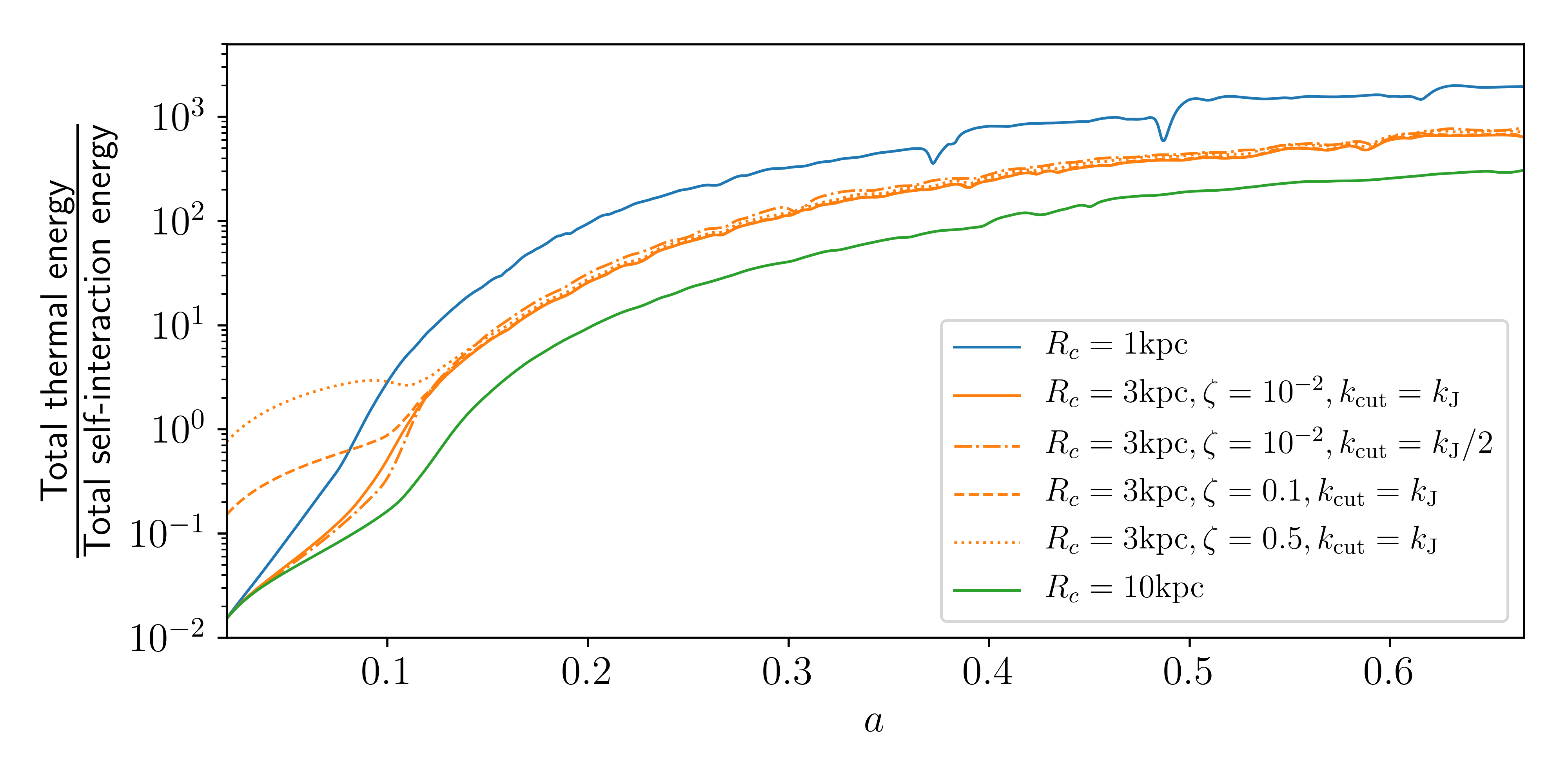}

    \caption{The evolution of the ratio of the total thermal energy and total self-interaction energy for the simulations run.}
\label{fig:U_over_U_SI}
\end{figure}

\begin{figure*}
    \centering
    \subfigure{\includegraphics[width=0.33\linewidth]{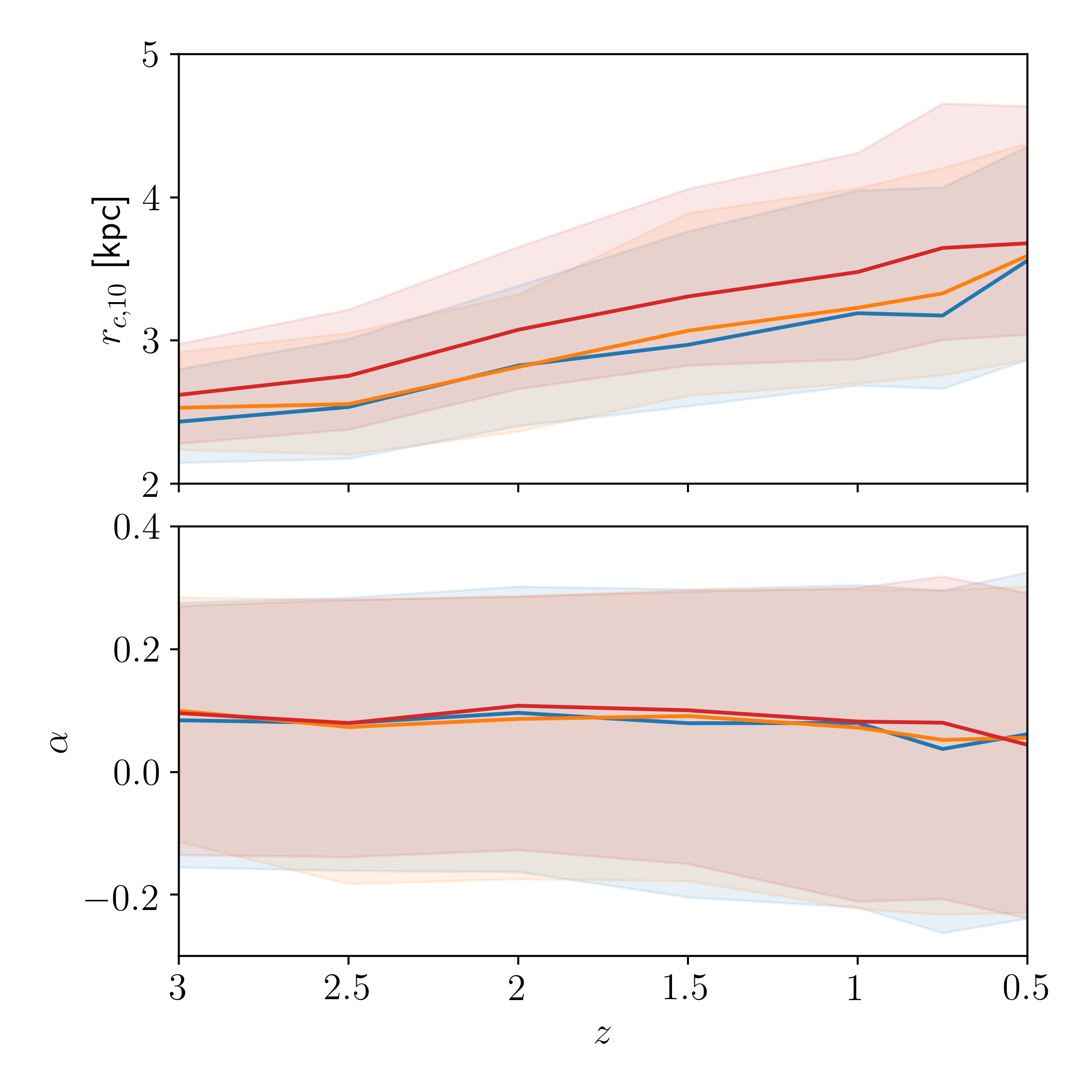}}
    \subfigure{\includegraphics[width=0.33\linewidth]{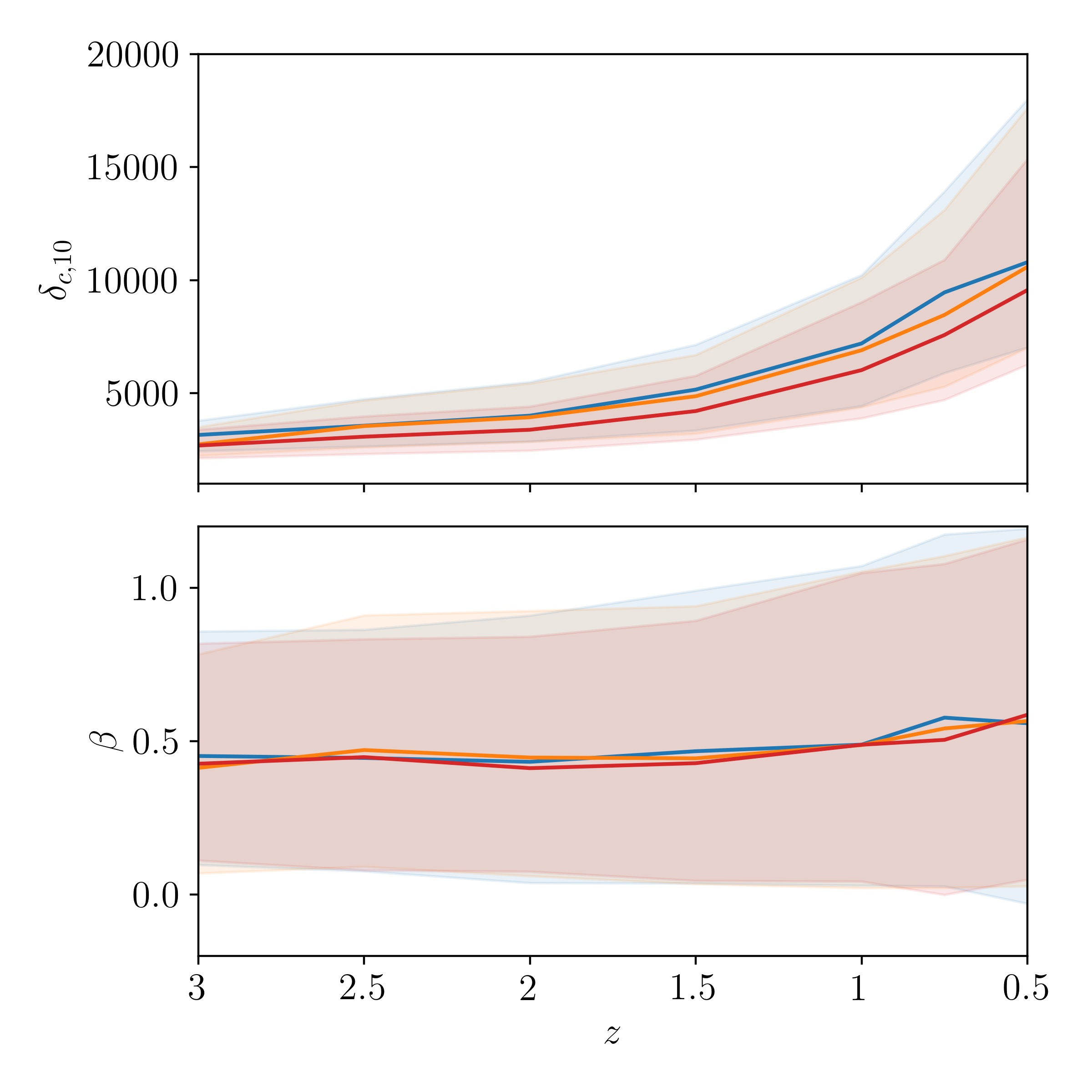}}
    \subfigure{\includegraphics[width=0.33\linewidth]{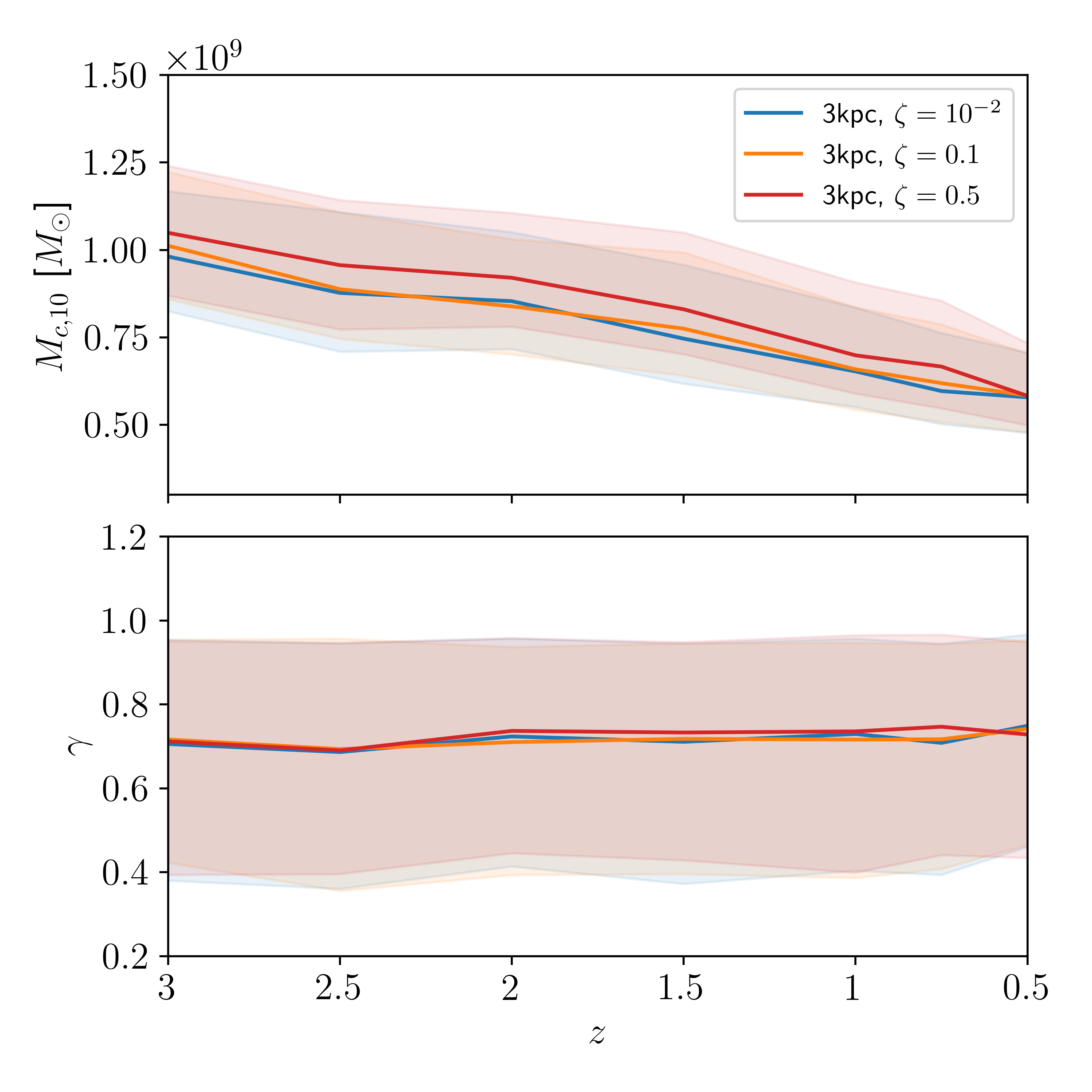}}
    \subfigure{\includegraphics[width=0.33\linewidth]{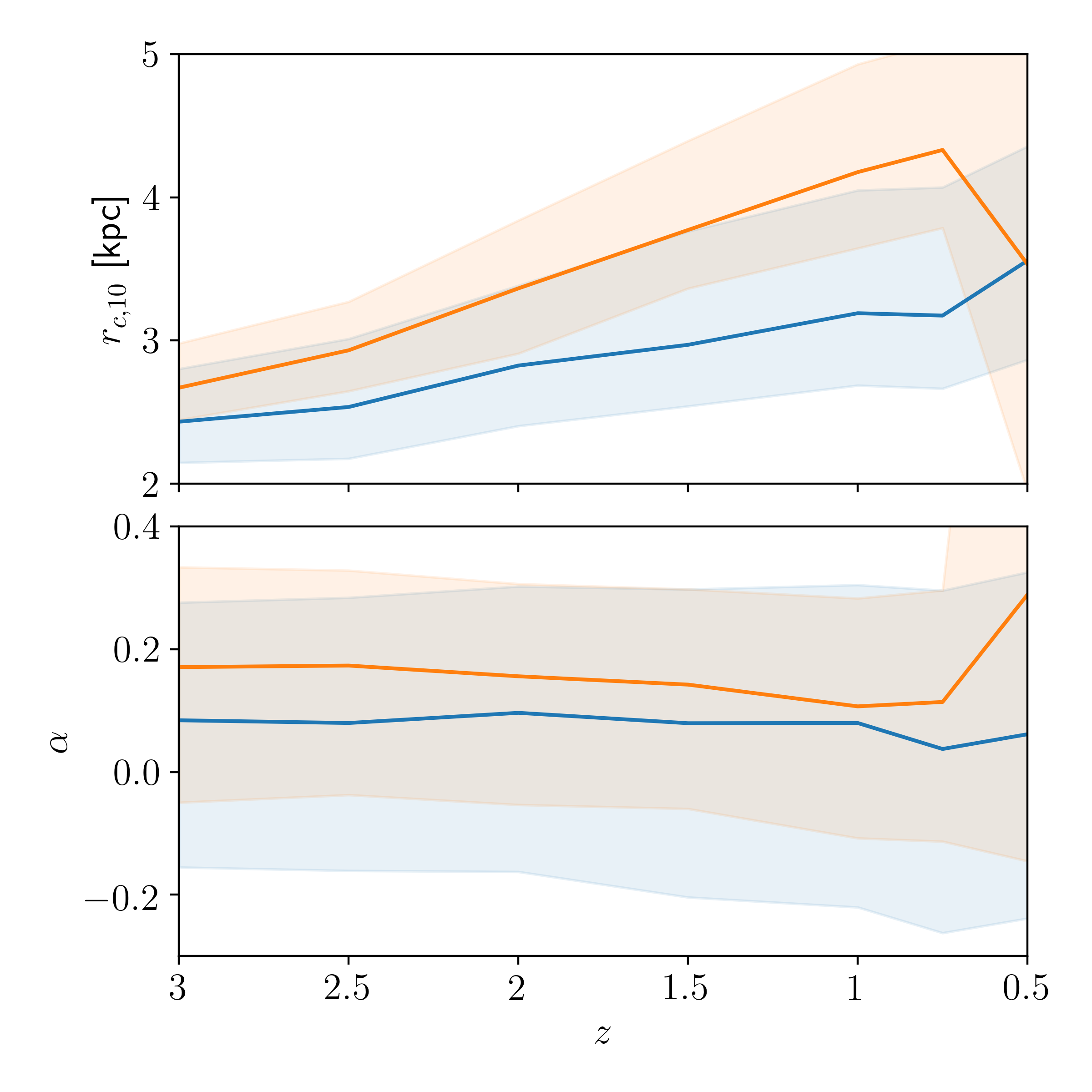}}
    \subfigure{\includegraphics[width=0.33\linewidth]{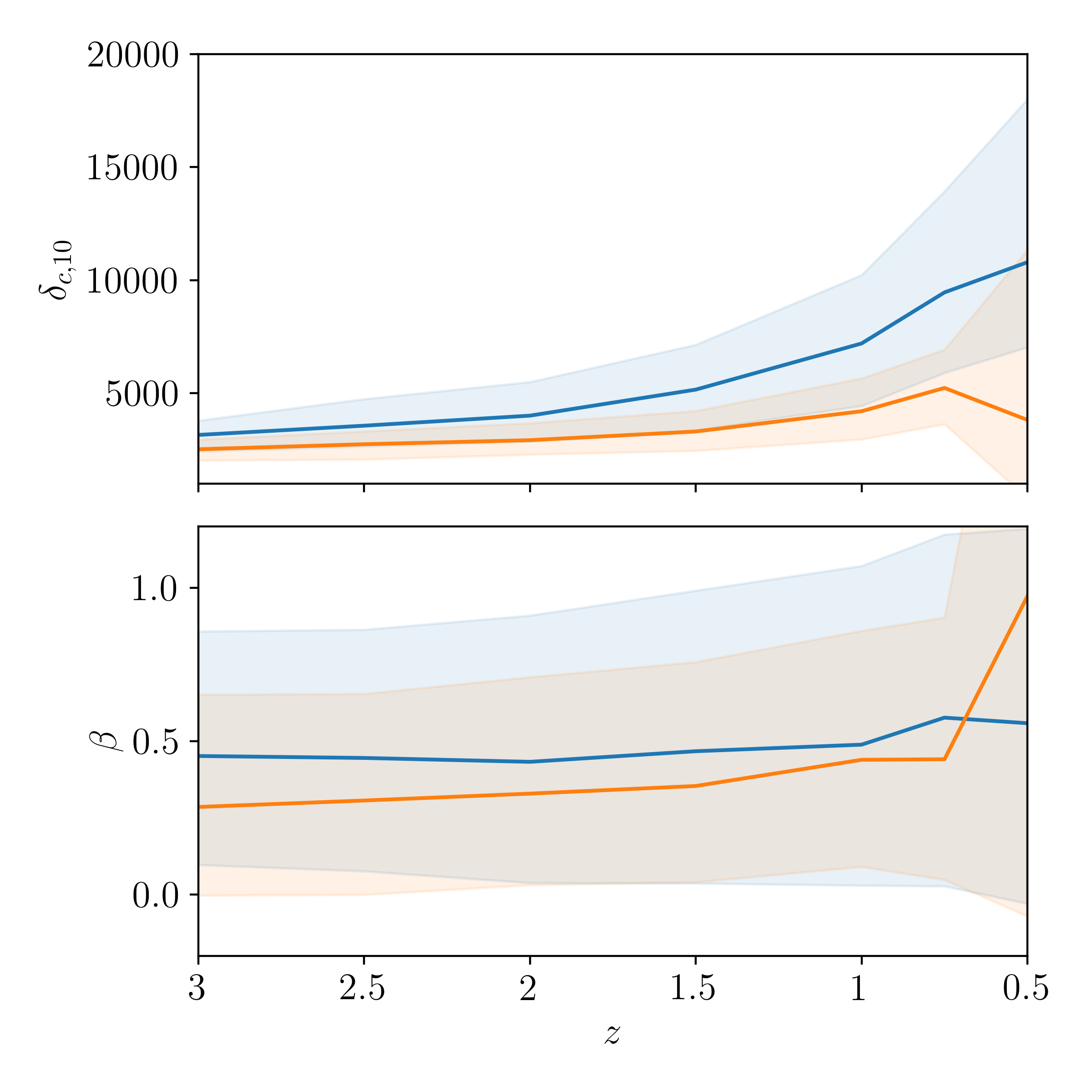}}
    \subfigure{\includegraphics[width=0.33\linewidth]{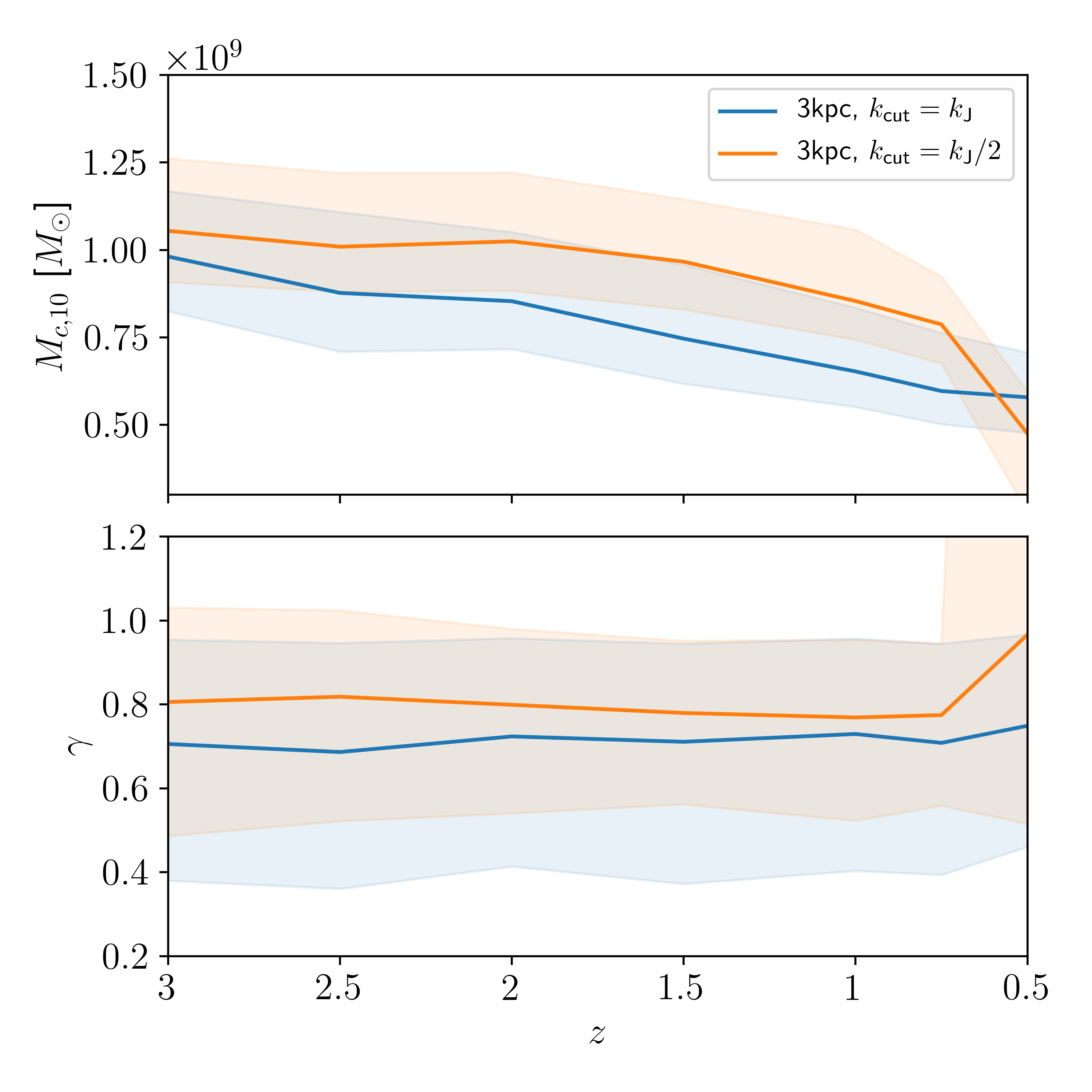}}

    \caption{The scaling functions at several redshifts for the core radii $r_c$ (\textit{left}), the core density $\delta_c$ (\textit{middle}), and the core mass $M_c$ (\textit{right}) for SIBEC-DM halos with $R_c=3\text{kpc}$ and different initial $\zeta = P/P_{\text{SI}}$ (\textit{upper}), and different initial power cut-offs (\textit{lower}). The shaded areas give the 1st and 3rd quartiles for the scaling parameters obtained from the Theil-Sen regression.}
\label{fig:Rc_3kpc_ic_comparisons}
\end{figure*}

The simulations are insensitive to the initial ratio $P/P_{\text{SI}}$ as long as it is relatively small, as shown in figures~\ref{fig:U_over_U_SI} and \ref{fig:Rc_3kpc_ic_comparisons}. As soon as structure begins to form, the thermal energy increases due to the conversion of kinetic energy to internal energy. A small amount of initial thermal energy therefore matters little. Nevertheless, despite the effective thermal energy dominating the DM-halo cores, the scale of the self-interaction determines the characteristic size of the cores.

Another interesting feature is the dependence of $\delta_c$ on the halo mass. The Burkert profile fitted to observational data prefers a negative slope, meaning massive halos are less dense. In $N$-body simulations of CDM, considering instead the characteristic over-density $\delta_s$ of the halo in eq. \eqref{eq:NFW_profile}, such mass dependence arises, in very simple terms, because $\delta_s$ is proportional to the mean density at the time the halo forms, with low-mass halos generally collapsing at a higher redshift, when the universe was denser \citep{Navarro1996}. We might therefore expect processes that transform the CDM halo cusps into cores, for instance baryonic feedback, to produce core densities that inherits the same halo mass dependence \citep{Ogiya2014}.
SIBEC-DM halos prefer instead a positive slope $\beta\approx 0.5$ at $z=0.5$, such that more massive halos are also denser. Furthermore, the SIBEC-DM halo core masses $M_c$ scales approximately as $M_c \sim M_{200}^{\gamma}$, with $\gamma \approx 0.75$ at $z=0.5$. The scaling of both $M_c$ and $\delta_c$ can be qualitatively understood under the assumption of velocity dispersion tracing outside the core, i.e. that the circular orbital velocity is nearly constant \citep{Chavanis2019a,Chavanis2019b,Padilla2021,Dawoodbhoy2021},
\begin{equation}
    v_c^2 = \frac{GM_c}{r_c} \approx v_{200}^2 = \frac{GM_{200}}{R_{200}},
\end{equation}
which we indeed see to be approximately true in figure~\ref{fig:halo_energy_contributions}, since $v^2 = -W/\rho$. Inserting that $r_c$ is constant, and $M_{200} \sim r_{200}^3$ gives
\begin{equation}
    M_c \sim M_{200}^{2/3},
\end{equation}
and
\begin{equation}
    \delta_c \propto \frac{M_c}{r_c^3} \sim M_{200}^{2/3},
\end{equation}
which are in the neighborhood of what is obtained in the simulations. We can improve this scaling argument by including a small mass dependence in $r_c\sim M_{200}^{\alpha}$, which gives
\begin{equation}
\label{eq:Mc_scaling_with_rc_exponent}
    M_c \sim M_{200}^{\frac{2}{3(1-\alpha)}},
\end{equation}
\begin{equation}
\label{eq:deltac_scaling_with_rc_exponent}
    \delta_c \sim M_{200}^{\frac{2}{3(1-\alpha)}-3\alpha}.
\end{equation}
Inserting $\alpha=0.05$ gives $M_c \sim M_{200}^{0.7}$ and $\delta_c \sim M_{200}^{0.55}$, while $\alpha=0.1$ gives $M_c \sim M_{200}^{0.74}$ and $\delta_c \sim M_{200}^{0.44}$.
For comparison, independent FDM simulations, using the full NLSE or the Madelung equations, find varying exponents, with $M_c \sim M_{200}^{1/3}$ \citep{Schive2014,Schive2014b,Veltmaat2018}, $M_c \sim M_{200}^{5/9}$ \citep{Mocz2017,Mina2020b}, or $M_c \sim M_{200}^{0.6}$ \citep{Nori2021}, all of which are less steep than what we find for SIBEC-DM. Furthermore, as expected from hydrostatic equilibrium, the size of FDM halo cores have been found to be decreasing with the virial mass \citep{Chan2022}, as opposed to slightly increasing for SIBEC-DM.

Generally, we find the trends in $r_c$, $\delta_c$, and $M_c$ for the values of $R_c$ and initial conditions tested in this paper to be in conflict with the SPARC dataset and Milky Way dSphs, although their slopes are in less tension with observation than FDM. These trends are mostly insensitive to the initial ratio $P/P_{\text{SI}}\ll 1$, but they are dependent on the cut-off scale in the initial matter power spectrum (when it is not scaled along with $R_c$ to match the Jeans' length), as shown in figure~\ref{fig:Rc_3kpc_ic_comparisons}. By moving the initial cut-off to larger scales, such that there is less initial power at small scales, the scaling exponents shift towards those inferred from the SPARC dataset and the Milky Way dSphs. On the flip-side, the prefactors shift away from the observations. For example, the total ratio $P/P_{\text{SI}}$ increases as $k_{\text{cut}}$ is lowered, meaning there is more thermal pressure, causing halos to have larger cores and lower central densities. However, as we saw in figure~\ref{fig:simulation_fitted_parameters}, decreasing the interaction strength, or equivalently, the core radius $R_c$, hardly affects the scaling exponents, but moves the prefactors in the desired direction. Based on this observation, one could imagine that a sufficiently small $R_c$ and strong cut-off might create a population of halos that alleviates to some degree the current tension with the observed trends. According to eqs. \eqref{eq:Mc_scaling_with_rc_exponent} and \eqref{eq:deltac_scaling_with_rc_exponent}, a stronger mass dependence in the core radius $r_c\sim M_{200}^{\alpha}$ will bring the SIBEC trends towards the observed ones, and can come about by having a larger thermal Jeans' length for massive halos, i.e. that more massive halos have a larger ratio of thermal energy to self-interaction energy, $U/U_{\text{SI}}$, due to e.g. multiple mergers.
However, caution is called for in making such a statement. For instance, the range of halo masses represented in the simulations is rather small, and baryonic physics are not at all included in the simulations, but are known to affect the distribution of DM in halos. Furthermore, the scaling function parameters have some time-dependence, primarily the prefactors, and our simulations were only run until $z=0.5$, whereas the observed galaxies and dSphs are at $z\approx 0$. This is particularly true for the Rc3-b run with $k_{\text{cut}}=k_{\text{J}}/2$, which has a factor $1/4$ fewer halos compared to $k_{\text{cut}}=k_{\text{J}}$, and is therefore more sensitive to disruptions in parts of the simulated halo population due to e.g. merger events. We see this in figure~\ref{fig:Rc_3kpc_ic_comparisons} at $z=0.5$, where the scaling parameters suddenly change, and the 1st and 3rd quartile-region becomes very large, signaling that the scaling relations generally do not provide a good fit to the halo population at that point.

\section{Conclusion}
\label{sec:conclusions}
In this paper a hydrodynamic approximation of the NLSE that includes the de Broglie-scale dynamics as an effective thermal energy was used to simulate the formation of structure of SIBEC-DM in a fully 3D cosmological setting. The advantage of such an approach is that the dynamics on the de Broglie-scale need not be resolved on the numerical grid, making the task of simulating SIBEC-DM simpler and computationally tractable. This is of particular importance for when the de Broglie wavelength is much smaller than the self-interaction Jeans' scale that we are interested in. On the other hand, this approach introduces an additional equation for the effective thermal energy must be evolved through time, and information of the underlying wavefunction is lost.

The self-interacting scalar field halos that are formed have NFW-like envelopes and cored interiors, and are well-described by a cored NFW profile such as eq. \eqref{eq:NFWc_profile} or the Burkert profile. The halo core radii are on the order of the hydrostatic core radius $R_c$, which is independent of the halo mass. Despite the self-interactions determining the scale of the cores, the effective thermal energy ends up dominating over the self-interactions throughout the halos due to significant heating during collapse, even in the cores. This is in contrast to previous 1D simulations that found the SIBEC-DM cores to not experience much heating and the thermal energy to fall sharply below the self-interaction energy inside the cores \citep{Dawoodbhoy2021, Shapiro2021}. This difference is due to the lack of mixing in spherically symmetric 1D simulations, whereas in 3D simulations matter collapses onto the halos in clumps rather than shells, and causes the outer shock-heated layers to be mixed with the cores.

Scaling relations for the core radii $r_c$, core densities $\delta_c$, and core masses $M_c$ as functions of the total halo mass $M_{200}$ were fitted to the simulated halo populations, which largely agrees with hydrostatic considerations of the halo cores where $r_c$ is nearly constant, and velocity dispersion tracing in the halo envelope, $v_c \approx v_{200}$. However, these trends do not agree with those obtained by fitting the Burkert profile to nearby galaxies in the SPARC dataset and the classical Milky Way dSphs. This poses an issue for SIBEC-DM models with $R_c \gtrsim 1\text{kpc}$ and a largely CDM-like matter power spectrum at late times \citep{Harko2011,Harko2012,Velteen2012,Freitas2013,Bettoni2014,Freitas2015,Hartman2022}, as was used in our simulations. For SIBEC-DM given by the field Largangian eq. \eqref{eq:complex_interacting_klein_gordon}, on the other hand, the self-interaction is constrained to $R_c < 1\text{kpc}$, otherwise an early radiation-like period and a large comoving Jeans' length washes out too much structure to be consistent with observations \citep{Shapiro2021,Hartman2022}. Halo masses affected in this way are given by eq. \eqref{eq:M_cut}. In fact, using observational constraints on FDM as a proxy for SIBEC-DM by matching their transfer function cut-offs and HMFs suggest the SIBEC-DM self-interaction to be as low as $R_c \sim 10\text{pc}$ \citep{Shapiro2021}. We were unable to probe such small values for $R_c$ since the large gap between the halo cores and the cut-off scale requires both a large simulation box and very high spatial resolution, though we find that simultaneously decreasing $R_c$ and increasing $k_{\text{cut}}$ might alleviate the tension in the scaling relations. It should be noted that our SIBEC-DM-only simulations do provide a better agreement with the slopes in observed scaling relations than FDM. In particular, FDM simulations generally find $M_c\sim M_{200}^{\gamma}$ with $1/3<\gamma<0.6$, while we find $\gamma\approx 0.75$, which is closer to the observed $\gamma\approx 1.1$. Additionally, FDM has a core radius that $r_c$ decreases with the halo mass, while we find a slightly increasing trend, although not as steep as found in the SPARC dataset and the Milky Way dSphs.

The main results of the present work is summarized as follows:
\begin{itemize}
\setlength\itemsep{0.5em}
    \item Using a hydrodynamic approximation and fully 3D cosmological simulations, SIBEC-DM halos are found to have NFW-like envelopes with central cores of radius $r_c$. The density profiles are therefore well-fitted by cored NFW profiles such as eq. \eqref{eq:NFWc_profile} and the Burkert profile, eq. \eqref{eq:Burkert_profile}.
    \item The SIBEC-DM cores are largely independent of the halo virial mass and central density, as predicted by hydrostatic equilibrium.
    \item Despite the self-interaction energy initially dominating the internal energy and fluid pressure of the SIBEC-DM, as well as determining the general scale of the SIBEC-DM cores, the effective thermal energy ends up dominating throughout the halo. This result is insensitive to the initial smallness of the thermal energy relative to the self-interaction energy, since the final thermal energy comes from heating as SIBEC-DM collapses and forms structure. 
    \item The SIBEC-DM halo core densities $\delta_c$ and core masses $M_c$ are found to scale with the virial mass $M_{200}$ as $\delta_c \sim M_{200}^{0.5}$ and $M_c \sim M_{200}^{0.75}$. This result is insensitive to changes to $R_c$ that is matched by a corresponding change in the cut-off scale,  $k_{\text{cut}}\sim 1/R_c$. Velocity tracing in the halo envelope and a nearly constant core radius predict these relations to be $\delta_c \sim M_c \sim M_{200}^{2/3}$. Including a small mass dependence in $r_c$ improves the agreement with the simulations, e.g. $r_c\sim M_{200}^{0.05}$ gives $M_c \sim M_{200}^{0.7}$ and $\delta_c \sim M_{200}^{0.55}$.
    \item The scaling relations for $r_c(M_{200})$, $\delta_c(M_{200})$, and $M_c(M_{200})$, eqs. \eqref{eq:rc_fitting_func}, \eqref{eq:deltac_fitting_func}, and \eqref{eq:Mc_fitting_func}, assuming the Burkert profile for the SIBEC-DM halos, generally do not agree with trends of the classical Milky Way dSphs and nearby galaxies in the SPARC dataset for the SIBEC-DM parameters tested in this work. However, there is some time-dependence in the scaling relations in our simulations (which were only run to redshift $z=0.5$), the prefactors more so than the exponents, and there is some dependence on the cut-off scale in the initial matter power spectrum used to generate the initial conditions. Nevertheless, the slopes of the scaling relations in our SIBEC-DM-only simulation are in better agreement with observations than in FDM-only simulations, and we find that they can be shifted towards the observed ones by making the dependence of $r_c$ on $M_{200}$ steeper, which can come about if massive halos have more thermal energy.
\end{itemize}

In future work, larger high-resolution simulations should be carried out to further investigate SIBEC-DM, as our current simulations have a limited box size and halo population. This is particularly relevant for probing SIBEC-DM with more realistic initial conditions and self-interactions that are consistent with the HMF \citep{Shapiro2021} and large-scale observables \citep{Hartman2022}. It will be important to test the validity and accuracy of the smoothing procedure and the resulting hydrodynamic approximation used in this work, especially if it continues to be used to test SIBEC-DM models. \citet{Dawoodbhoy2021} showed analytically for a few idealized 1D cases that smoothing the exact solutions of the NLSE gives the same large-scale fluid properties, such as pressure and density, as the smoothed phase space distribution function. These tests should be extended to more complicated scenarios, such as self-gravitation and the formation of shock fronts, to check if, for instance, the heating of the effective thermal energy is accurately captured by the hydrodynamic approximation, or if the skewlessness of $\mathcal{F}$ is a valid assumption, although this will require using numerical simulations.

\begin{acknowledgements}
We thank the Research Council of Norway for their support, and Matteo Nori and Marco Baldi for discussions on simulations of ultra-light DM. We also extend our gratitude to Taha Dawoodbhoy for his feedback on this work. Computations were performed on resources provided by UNINETT Sigma2 -- the National Infrastructure for High Performance Computing and Data Storage in Norway.
\end{acknowledgements}


\begin{thebibliography}{86}
\expandafter\ifx\csname natexlab\endcsname\relax\def\natexlab#1{#1}\fi

\bibitem[{Arbey {et~al.}(2002)Arbey, Lesgourgues, \& Salati}]{Arbey2002}
Arbey, A., Lesgourgues, J., \& Salati, P. 2002, Physical Review D, 65, 083514,
  publisher: American Physical Society

\bibitem[{Arbey {et~al.}(2003)Arbey, Lesgourgues, \& Salati}]{Arbey2003}
Arbey, A., Lesgourgues, J., \& Salati, P. 2003, Physical Review D, 68, 023511,
  publisher: American Physical Society

\bibitem[{Berezhiani {et~al.}(2019)Berezhiani, Elder, \&
  Khoury}]{Berezhiani2019}
Berezhiani, L., Elder, B., \& Khoury, J. 2019, Journal of Cosmology and
  Astroparticle Physics, 2019, 074

\bibitem[{Berezhiani \& Khoury(2015)}]{Berezhiani2015}
Berezhiani, L. \& Khoury, J. 2015, Phys. Rev. D, 92, 103510

\bibitem[{Bettoni {et~al.}(2014)Bettoni, Colombo, \& Liberati}]{Bettoni2014}
Bettoni, D., Colombo, M., \& Liberati, S. 2014, Journal of Cosmology and
  Astroparticle Physics, 2014, 004, publisher: IOP Publishing

\bibitem[{Bullock \& Boylan-Kolchin(2017)}]{Bullock2017}
Bullock, J.~S. \& Boylan-Kolchin, M. 2017, Annual Review of Astronomy and
  Astrophysics, 55, 343

\bibitem[{Burkert(1995)}]{Burkert1995}
Burkert, A. 1995, The Astrophysical Journal, 447, L25, {ADS} Bibcode:
  1995ApJ...447L..25B

\bibitem[{Böhmer \& Harko(2007)}]{Bohmer2007}
Böhmer, C.~G. \& Harko, T. 2007, Journal of Cosmology and Astroparticle
  Physics, 2007, 025, publisher: IOP Publishing

\bibitem[{Chan {et~al.}(2022)Chan, Ferreira, May, Hayashi, \& Chiba}]{Chan2022}
Chan, H. Y.~J., Ferreira, E. G.~M., May, S., Hayashi, K., \& Chiba, M. 2022,
  Monthly Notices of the Royal Astronomical Society, 511, 943

\bibitem[{Chavanis(2011)}]{Chavanis2011}
Chavanis, P.-H. 2011, Physical Review D, 84, 043531, publisher: American
  Physical Society

\bibitem[{Chavanis(2019{\natexlab{a}})}]{Chavanis2019a}
Chavanis, P.-H. 2019{\natexlab{a}}, Physical Review D, 100, 123506, publisher:
  American Physical Society

\bibitem[{Chavanis(2019{\natexlab{b}})}]{Chavanis2019b}
Chavanis, P.-H. 2019{\natexlab{b}}, Physical Review D, 100, 083022, publisher:
  American Physical Society

\bibitem[{Chavanis \& Delfini(2011)}]{Chavanis2011b}
Chavanis, P.-H. \& Delfini, L. 2011, Physical Review D, 84, 043532, publisher:
  American Physical Society

\bibitem[{Crăciun \& Harko(2020)}]{Craciun2020}
Crăciun, M. \& Harko, T. 2020, arXiv:2007.12222 [astro-ph, physics:gr-qc,
  physics:hep-th], arXiv: 2007.12222

\bibitem[{Cyburt {et~al.}(2016)Cyburt, Fields, Olive, \& Yeh}]{Cyburt2016}
Cyburt, R.~H., Fields, B.~D., Olive, K.~A., \& Yeh, T.-H. 2016, Reviews of
  Modern Physics, 88, 015004, publisher: American Physical Society

\bibitem[{Davis {et~al.}(1985)Davis, Efstathiou, Frenk, \& White}]{Davis1985}
Davis, M., Efstathiou, G., Frenk, C.~S., \& White, S. D.~M. 1985, The
  Astrophysical Journal, 292, 371

\bibitem[{Dawoodbhoy {et~al.}(2021)Dawoodbhoy, Shapiro, \&
  Rindler-Daller}]{Dawoodbhoy2021}
Dawoodbhoy, T., Shapiro, P.~R., \& Rindler-Daller, T. 2021, Monthly Notices of
  the Royal Astronomical Society, 506, 2418

\bibitem[{de~Freitas \& Velten(2015)}]{Freitas2015}
de~Freitas, R.~C. \& Velten, H. 2015, The European Physical Journal C, 75, 597

\bibitem[{Del~Popolo \& Le~Delliou(2017)}]{DelPopolo2017}
Del~Popolo, A. \& Le~Delliou, M. 2017, Galaxies, 5

\bibitem[{Dine \& Fischler(1983)}]{Dine1983}
Dine, M. \& Fischler, W. 1983, Physics Letters B, 120, 137

\bibitem[{Einasto(1965)}]{Einasto1965}
Einasto, J. 1965, Trudy Astrofizicheskogo Instituta Alma-Ata, 5, 87, {ADS}
  Bibcode: 1965TrAlm...5...87E

\bibitem[{Ferreira(2020)}]{Ferreira2020}
Ferreira, E. G.~M. 2020, arXiv:2005.03254 [astro-ph, physics:cond-mat,
  physics:gr-qc, physics:hep-th], arXiv: 2005.03254

\bibitem[{Ferreira {et~al.}(2019)Ferreira, Franzmann, Khoury, \&
  Brandenberger}]{Ferreira2019}
Ferreira, E. G.~M., Franzmann, G., Khoury, J., \& Brandenberger, R. 2019,
  Journal of Cosmology and Astroparticle Physics, 2019, 027, arXiv: 1810.09474

\bibitem[{Freitas \& Gonçalves(2013)}]{Freitas2013}
Freitas, R.~C. \& Gonçalves, S. V.~B. 2013, Journal of Cosmology and
  Astroparticle Physics, 2013, 049, publisher: IOP Publishing

\bibitem[{Gill {et~al.}(2004)Gill, Knebe, \& Gibson}]{Gill2004}
Gill, S. P.~D., Knebe, A., \& Gibson, B.~K. 2004, Monthly Notices of the Royal
  Astronomical Society, 351, 399

\bibitem[{Goodman(2000)}]{Goodman2000}
Goodman, J. 2000, New Astronomy, 5, 103

\bibitem[{Hahn \& Abel(2011)}]{Hahn2011}
Hahn, O. \& Abel, T. 2011, Monthly Notices of the Royal Astronomical Society,
  415, 2101

\bibitem[{Harko(2011)}]{Harko2011}
Harko, T. 2011, Physical Review D, 83, 123515

\bibitem[{Harko \& Mocanu(2012)}]{Harko2012}
Harko, T. \& Mocanu, G. 2012, Physical Review D, 85, 084012

\bibitem[{Hartman {et~al.}(2022)Hartman, Winther, \& Mota}]{Hartman2022}
Hartman, S. T.~H., Winther, H.~A., \& Mota, D.~F. 2022, Journal of Cosmology
  and Astroparticle Physics, 2022, 005, publisher: {IOP} Publishing

\bibitem[{Hu {et~al.}(2000)Hu, Barkana, \& Gruzinov}]{Hu2000}
Hu, W., Barkana, R., \& Gruzinov, A. 2000, Phys. Rev. Lett., 85, 1158

\bibitem[{Huang(1987)}]{Huang1987}
Huang, K. 1987, Statistical Mechanics, 2nd edn. (John Wiley \& Sons)

\bibitem[{Hui {et~al.}(2017)Hui, Ostriker, Tremaine, \& Witten}]{Hui2017}
Hui, L., Ostriker, J.~P., Tremaine, S., \& Witten, E. 2017, Physical Review D,
  95, 043541

\bibitem[{Husimi(1940)}]{Husimi1940}
Husimi, K. 1940, Proceedings of the Physico-Mathematical Society of Japan. 3rd
  Series, 22, 264

\bibitem[{Khoury(2016)}]{Khoury2016}
Khoury, J. 2016, Phys. Rev. D, 93, 103533

\bibitem[{Knollmann \& Knebe(2009)}]{Knollmann2009}
Knollmann, S.~R. \& Knebe, A. 2009, The Astrophysical Journal Supplement
  Series, 182, 608, publisher: American Astronomical Society

\bibitem[{Lancaster {et~al.}(2020)Lancaster, Giovanetti, Mocz, Kahn, Lisanti,
  \& Spergel}]{Lancaster2020}
Lancaster, L., Giovanetti, C., Mocz, P., {et~al.} 2020, Journal of Cosmology
  and Astroparticle Physics, 2020, 001, publisher: IOP Publishing

\bibitem[{Lee \& Koh(1996)}]{Lee1996}
Lee, J.-w. \& Koh, I.-g. 1996, Physical Review D, 53, 2236, publisher: American
  Physical Society

\bibitem[{Lelli {et~al.}(2016)Lelli, McGaugh, \& Schombert}]{Lelli2016}
Lelli, F., McGaugh, S.~S., \& Schombert, J.~M. 2016, The Astronomical Journal,
  152, 157, publisher: American Astronomical Society

\bibitem[{Li {et~al.}(2014)Li, Rindler-Daller, \& Shapiro}]{Li2014}
Li, B., Rindler-Daller, T., \& Shapiro, P.~R. 2014, Physical Review D, 89,
  083536, publisher: American Physical Society

\bibitem[{Li {et~al.}(2020)Li, Lelli, {McGaugh}, \& Schombert}]{Li2020}
Li, P., Lelli, F., {McGaugh}, S., \& Schombert, J. 2020, The Astrophysical
  Journal Supplement Series, 247, 31, publisher: American Astronomical Society

\bibitem[{Madelung(1926)}]{Madelung1926}
Madelung, E. 1926, Naturwissenschaften, 14, 1004

\bibitem[{Marsh(2016)}]{Marsh2016}
Marsh, D. J.~E. 2016, Physics Reports, 643, 1

\bibitem[{Martel \& Shapiro(1998)}]{Martel1998}
Martel, H. \& Shapiro, P.~R. 1998, Mon. Not. R. Astron. Soc, 297, 467

\bibitem[{Matos \& Arturo Ureña-López(2001)}]{Matos2001}
Matos, T. \& Arturo Ureña-López, L. 2001, Physical Review D, 63, 063506,
  publisher: American Physical Society

\bibitem[{Matos {et~al.}(2000)Matos, Guzmán, \& Ureña-López}]{Matos2000}
Matos, T., Guzmán, F.~S., \& Ureña-López, L.~A. 2000, Classical and Quantum
  Gravity, 17, 1707, publisher: IOP Publishing

\bibitem[{May \& Springel(2021)}]{May2021}
May, S. \& Springel, V. 2021, arXiv:2101.01828 [astro-ph, physics:gr-qc],
  arXiv: 2101.01828

\bibitem[{Mina {et~al.}(2020{\natexlab{a}})Mina, Mota, \& Winther}]{Mina2020}
Mina, M., Mota, D.~F., \& Winther, H.~A. 2020{\natexlab{a}}, Astronomy \&
  Astrophysics, 641, A107, publisher: EDP Sciences

\bibitem[{Mina {et~al.}(2020{\natexlab{b}})Mina, Mota, \& Winther}]{Mina2020b}
Mina, M., Mota, D.~F., \& Winther, H.~A. 2020{\natexlab{b}}, arXiv:2007.04119
  [astro-ph, physics:gr-qc], arXiv: 2007.04119

\bibitem[{Mocz {et~al.}(2018)Mocz, Lancaster, Fialkov, Becerra, \&
  Chavanis}]{Mocz2018}
Mocz, P., Lancaster, L., Fialkov, A., Becerra, F., \& Chavanis, P.-H. 2018,
  Physical Review D, 97, 083519

\bibitem[{Mocz {et~al.}(2017)Mocz, Vogelsberger, Robles, Zavala,
  Boylan-Kolchin, Fialkov, \& Hernquist}]{Mocz2017}
Mocz, P., Vogelsberger, M., Robles, V.~H., {et~al.} 2017, Monthly Notices of
  the Royal Astronomical Society, 471, 4559

\bibitem[{Navarro {et~al.}(1996)Navarro, Frenk, \& White}]{Navarro1996}
Navarro, J.~F., Frenk, C.~S., \& White, S. D.~M. 1996, The Astrophysical
  Journal, 462, 563

\bibitem[{Navarro {et~al.}(1997)Navarro, Frenk, \& White}]{Navarro1997}
Navarro, J.~F., Frenk, C.~S., \& White, S. D.~M. 1997, The Astrophysical
  Journal, 490, 493, publisher: American Astronomical Society

\bibitem[{Nori \& Baldi(2018)}]{Nori2018}
Nori, M. \& Baldi, M. 2018, Monthly Notices of the Royal Astronomical Society,
  478, 3935, publisher: Oxford Academic

\bibitem[{Nori \& Baldi(2021)}]{Nori2021}
Nori, M. \& Baldi, M. 2021, Monthly Notices of the Royal Astronomical Society,
  501, 1539

\bibitem[{Ogiya {et~al.}(2014)Ogiya, Mori, Ishiyama, \& Burkert}]{Ogiya2014}
Ogiya, G., Mori, M., Ishiyama, T., \& Burkert, A. 2014, Monthly Notices of the
  Royal Astronomical Society: Letters, 440, L71

\bibitem[{Padilla {et~al.}(2019)Padilla, Rindler-Daller, Shapiro, Matos, \&
  Vázquez}]{Padilla2021}
Padilla, L.~E., Rindler-Daller, T., Shapiro, P.~R., Matos, T., \& Vázquez,
  J.~A. 2019, Physical Review D, 103, 063012, publisher: American Physical
  Society

\bibitem[{Peebles(2000)}]{Peebles2000}
Peebles, P. J.~E. 2000, The Astrophysical Journal, 534, L127, publisher: IOP
  Publishing

\bibitem[{Percival {et~al.}(2001)Percival, Baugh, Bland-Hawthorn, Bridges,
  Cannon, Cole, Colless, Collins, Couch, Dalton, De~Propris, Driver,
  Efstathiou, Ellis, Frenk, Glazebrook, Jackson, Lahav, Lewis, Lumsden, Maddox,
  Moody, Norberg, Peacock, Peterson, Sutherland, \& Taylor}]{Percival2001}
Percival, W.~J., Baugh, C.~M., Bland-Hawthorn, J., {et~al.} 2001, Monthly
  Notices of the Royal Astronomical Society, 327, 1297, publisher: Oxford
  Academic

\bibitem[{Pitaevskii \& Stringari(2016)}]{Pitaevskii2016}
Pitaevskii, L.~P. \& Stringari, S. 2016, Bose-Einstein Condensation and
  Superfluidity (Great Clarendon Street, Oxford, United Kingdom: Oxford
  University Press)

\bibitem[{{Planck Collaboration} {et~al.}(2016){Planck Collaboration}, {Ade},
  {Aghanim}, {Arnaud}, {Ashdown}, {Aumont}, {Baccigalupi}, {Banday},
  {Barreiro}, {Bartlett}, {Bartolo}, {Battaner}, {Battye}, {Benabed},
  {Beno{\^\i}t}, {Benoit-L{\'e}vy}, {Bernard}, {Bersanelli}, {Bielewicz},
  {Bock}, {Bonaldi}, {Bonavera}, {Bond}, {Borrill}, {Bouchet}, {Boulanger},
  {Bucher}, {Burigana}, {Butler}, {Calabrese}, {Cardoso}, {Catalano},
  {Challinor}, {Chamballu}, {Chary}, {Chiang}, {Chluba}, {Christensen},
  {Church}, {Clements}, {Colombi}, {Colombo}, {Combet}, {Coulais}, {Crill},
  {Curto}, {Cuttaia}, {Danese}, {Davies}, {Davis}, {de Bernardis}, {de Rosa},
  {de Zotti}, {Delabrouille}, {D{\'e}sert}, {Di Valentino}, {Dickinson},
  {Diego}, {Dolag}, {Dole}, {Donzelli}, {Dor{\'e}}, {Douspis}, {Ducout},
  {Dunkley}, {Dupac}, {Efstathiou}, {Elsner}, {En{\ss}lin}, {Eriksen},
  {Farhang}, {Fergusson}, {Finelli}, {Forni}, {Frailis}, {Fraisse},
  {Franceschi}, {Frejsel}, {Galeotta}, {Galli}, {Ganga}, {Gauthier}, {Gerbino},
  {Ghosh}, {Giard}, {Giraud-H{\'e}raud}, {Giusarma}, {Gjerl{\o}w},
  {Gonz{\'a}lez-Nuevo}, {G{\'o}rski}, {Gratton}, {Gregorio}, {Gruppuso},
  {Gudmundsson}, {Hamann}, {Hansen}, {Hanson}, {Harrison}, {Helou},
  {Henrot-Versill{\'e}}, {Hern{\'a}ndez-Monteagudo}, {Herranz}, {Hildebrand t},
  {Hivon}, {Hobson}, {Holmes}, {Hornstrup}, {Hovest}, {Huang}, {Huffenberger},
  {Hurier}, {Jaffe}, {Jaffe}, {Jones}, {Juvela}, {Keih{\"a}nen}, {Keskitalo},
  {Kisner}, {Kneissl}, {Knoche}, {Knox}, {Kunz}, {Kurki-Suonio}, {Lagache},
  {L{\"a}hteenm{\"a}ki}, {Lamarre}, {Lasenby}, {Lattanzi}, {Lawrence}, {Leahy},
  {Leonardi}, {Lesgourgues}, {Levrier}, {Lewis}, {Liguori}, {Lilje},
  {Linden-V{\o}rnle}, {L{\'o}pez-Caniego}, {Lubin}, {Mac{\'\i}as-P{\'e}rez},
  {Maggio}, {Maino}, {Mandolesi}, {Mangilli}, {Marchini}, {Maris}, {Martin},
  {Martinelli}, {Mart{\'\i}nez-Gonz{\'a}lez}, {Masi}, {Matarrese}, {McGehee},
  {Meinhold}, {Melchiorri}, {Melin}, {Mendes}, {Mennella}, {Migliaccio},
  {Millea}, {Mitra}, {Miville-Desch{\^e}nes}, {Moneti}, {Montier}, {Morgante},
  {Mortlock}, {Moss}, {Munshi}, {Murphy}, {Naselsky}, {Nati}, {Natoli},
  {Netterfield}, {N{\o}rgaard-Nielsen}, {Noviello}, {Novikov}, {Novikov},
  {Oxborrow}, {Paci}, {Pagano}, {Pajot}, {Paladini}, {Paoletti}, {Partridge},
  {Pasian}, {Patanchon}, {Pearson}, {Perdereau}, {Perotto}, {Perrotta},
  {Pettorino}, {Piacentini}, {Piat}, {Pierpaoli}, {Pietrobon}, {Plaszczynski},
  {Pointecouteau}, {Polenta}, {Popa}, {Pratt}, {Pr{\'e}zeau}, {Prunet},
  {Puget}, {Rachen}, {Reach}, {Rebolo}, {Reinecke}, {Remazeilles}, {Renault},
  {Renzi}, {Ristorcelli}, {Rocha}, {Rosset}, {Rossetti}, {Roudier},
  {Rouill{\'e} d'Orfeuil}, {Rowan-Robinson}, {Rubi{\~n}o-Mart{\'\i}n},
  {Rusholme}, {Said}, {Salvatelli}, {Salvati}, {Sandri}, {Santos},
  {Savelainen}, {Savini}, {Scott}, {Seiffert}, {Serra}, {Shellard}, {Spencer},
  {Spinelli}, {Stolyarov}, {Stompor}, {Sudiwala}, {Sunyaev}, {Sutton},
  {Suur-Uski}, {Sygnet}, {Tauber}, {Terenzi}, {Toffolatti}, {Tomasi},
  {Tristram}, {Trombetti}, {Tucci}, {Tuovinen}, {T{\"u}rler}, {Umana},
  {Valenziano}, {Valiviita}, {Van Tent}, {Vielva}, {Villa}, {Wade}, {Wandelt},
  {Wehus}, {White}, {White}, {Wilkinson}, {Yvon}, {Zacchei}, \&
  {Zonca}}]{Planck2015}
{Planck Collaboration}, {Ade}, P.~A.~R., {Aghanim}, N., {et~al.} 2016, \aap,
  594, A13

\bibitem[{Preskill {et~al.}(1983)Preskill, Wise, \& Wilczek}]{Preskill1983}
Preskill, J., Wise, M.~B., \& Wilczek, F. 1983, Physics Letters B, 120, 127

\bibitem[{Riess {et~al.}(2016)Riess, Macri, Hoffmann, Scolnic, Casertano,
  Filippenko, Tucker, Reid, Jones, Silverman, Chornock, Challis, Yuan, Brown,
  \& Foley}]{Riess2016}
Riess, A.~G., Macri, L.~M., Hoffmann, S.~L., {et~al.} 2016, The Astrophysical
  Journal, 826, 56, publisher: American Astronomical Society

\bibitem[{Rindler-Daller \& Shapiro(2012)}]{Rindler-Daller2012}
Rindler-Daller, T. \& Shapiro, P.~R. 2012, Monthly Notices of the Royal
  Astronomical Society, 422, 135

\bibitem[{Rindler-Daller \& Shapiro(2014)}]{Rindler-Daller2014}
Rindler-Daller, T. \& Shapiro, P.~R. 2014, Modern Physics Letters A, 29,
  1430002, publisher: World Scientific Publishing Co.

\bibitem[{Robles {et~al.}(2019)Robles, Bullock, \& Boylan-Kolchin}]{Robles2019}
Robles, V.~H., Bullock, J.~S., \& Boylan-Kolchin, M. 2019, Monthly Notices of
  the Royal Astronomical Society, 483, 289

\bibitem[{Rogers \& Peiris(2021)}]{Rogers2021}
Rogers, K.~K. \& Peiris, H.~V. 2021, Physical Review Letters, 126, 071302,
  publisher: American Physical Society

\bibitem[{Salucci {et~al.}(2012)Salucci, Wilkinson, Walker, Gilmore, Grebel,
  Koch, Frigerio~Martins, \& Wyse}]{Salucci2012}
Salucci, P., Wilkinson, M.~I., Walker, M.~G., {et~al.} 2012, Monthly Notices of
  the Royal Astronomical Society, 420, 2034

\bibitem[{Schive {et~al.}(2014{\natexlab{a}})Schive, Chiueh, \&
  Broadhurst}]{Schive2014}
Schive, H.-Y., Chiueh, T., \& Broadhurst, T. 2014{\natexlab{a}}, Nature
  Physics, 10, 496

\bibitem[{Schive {et~al.}(2014{\natexlab{b}})Schive, Liao, Woo, Wong, Chiueh,
  Broadhurst, \& Hwang}]{Schive2014b}
Schive, H.-Y., Liao, M.-H., Woo, T.-P., {et~al.} 2014{\natexlab{b}}, Physical
  Review Letters, 113, 261302, publisher: American Physical Society

\bibitem[{Schwabe {et~al.}(2016)Schwabe, Niemeyer, \& Engels}]{Schwabe2016}
Schwabe, B., Niemeyer, J.~C., \& Engels, J.~F. 2016, Phys. Rev. D, 94, 043513

\bibitem[{Sen(1968)}]{Sen1968}
Sen, P.~K. 1968, Journal of the American Statistical Association, 63, 1379,
  publisher: Taylor \& Francis

\bibitem[{Shapiro {et~al.}(2021)Shapiro, Dawoodbhoy, \&
  Rindler-Daller}]{Shapiro2021}
Shapiro, P.~R., Dawoodbhoy, T., \& Rindler-Daller, T. 2021, Monthly Notices of
  the Royal Astronomical Society, 509, 145

\bibitem[{Skodje {et~al.}(1989)Skodje, Rohrs, \& VanBuskirk}]{Skodje1989}
Skodje, R.~T., Rohrs, H.~W., \& VanBuskirk, J. 1989, Physical Review A, 40,
  2894, publisher: American Physical Society

\bibitem[{Tegmark {et~al.}(2004)Tegmark, Blanton, Strauss, Hoyle, Schlegel,
  Scoccimarro, Vogeley, Weinberg, Zehavi, Berlind, Budavari, Connolly,
  Eisenstein, Finkbeiner, Frieman, Gunn, Hamilton, Hui, Jain, Johnston, Kent,
  Lin, Nakajima, Nichol, Ostriker, Pope, Scranton, Seljak, Sheth, Stebbins,
  Szalay, Szapudi, Verde, Xu, Annis, Bahcall, Brinkmann, Burles, Castander,
  Csabai, Loveday, Doi, Fukugita, III, Hennessy, Hogg, Ivezi{\'{c}}, Knapp,
  Lamb, Lee, Lupton, McKay, Kunszt, Munn, O'Connell, Peoples, Pier, Richmond,
  Rockosi, Schneider, Stoughton, Tucker, Berk, Yanny, \& and}]{Tegmark2004}
Tegmark, M., Blanton, M.~R., Strauss, M.~A., {et~al.} 2004, The Astrophysical
  Journal, 606, 702

\bibitem[{Teyssier(2002)}]{Teyssier2002}
Teyssier, R. 2002, Astronomy \& Astrophysics, 385, 337, number: 1 Publisher:
  EDP Sciences

\bibitem[{Theil(1950)}]{Theil1950}
Theil, H. 1950, Nederl. Akad. Wetensch., Proc., 53, 386

\bibitem[{Toro(2006)}]{Toro2006}
Toro, E. 2006, Applied Numerical Mathematics, 56, 1464

\bibitem[{Trujillo-Gomez {et~al.}(2011)Trujillo-Gomez, Klypin, Primack, \&
  Romanowsky}]{Trujillo-Gomez2011}
Trujillo-Gomez, S., Klypin, A., Primack, J., \& Romanowsky, A.~J. 2011, The
  Astrophysical Journal, 742, 16, publisher: American Astronomical Society

\bibitem[{Turk {et~al.}(2010)Turk, Smith, Oishi, Skory, Skillman, Abel, \&
  Norman}]{Turk2010}
Turk, M.~J., Smith, B.~D., Oishi, J.~S., {et~al.} 2010, The Astrophysical
  Journal Supplement Series, 192, 9, publisher: American Astronomical Society

\bibitem[{Velten \& Wamba(2012)}]{Velteen2012}
Velten, H. \& Wamba, E. 2012, Physics Letters B, 709, 1

\bibitem[{Veltmaat {et~al.}(2018)Veltmaat, Niemeyer, \& Schwabe}]{Veltmaat2018}
Veltmaat, J., Niemeyer, J.~C., \& Schwabe, B. 2018, Physical Review D, 98,
  043509

\bibitem[{Vogelsberger {et~al.}(2014)Vogelsberger, Genel, Springel, Torrey,
  Sijacki, Xu, Snyder, Bird, Nelson, \& Hernquist}]{Vogelsberger2014}
Vogelsberger, M., Genel, S., Springel, V., {et~al.} 2014, Nature, 509, 177,
  number: 7499 Publisher: Nature Publishing Group

\bibitem[{Weinberg {et~al.}(2015)Weinberg, Bullock, Governato, Naray, \&
  Peter}]{Weinberg2015}
Weinberg, D.~H., Bullock, J.~S., Governato, F., Naray, R. K.~d., \& Peter, A.
  H.~G. 2015, Proceedings of the National Academy of Sciences, 112, 12249,
  publisher: National Academy of Sciences Section: Colloquium Paper

\bibitem[{Widrow \& Kaiser(1993)}]{Widrow1993}
Widrow, L.~M. \& Kaiser, N. 1993, The Astrophysical Journal, 416, L71

\bibitem[{Zhang {et~al.}(2018)Zhang, Chan, Harko, Liang, \& Leung}]{Zhang2018}
Zhang, X., Chan, M.~H., Harko, T., Liang, S.-D., \& Leung, C.~S. 2018, The
  European Physical Journal C, 78, 346

\end{thebibliography}


\onecolumn
\appendix

\section{Fitted scaling relation parameters}
\label{app:fitted_scaling_relation_parameters}
The fitted parameters for the scaling relations of $r_c(M_{200})$, $\delta_c(M_{200})$, and $M_c(M_{200})$, i.e. eqs. \eqref{eq:rc_fitting_func}, \eqref{eq:deltac_fitting_func}, and \eqref{eq:Mc_fitting_func}, are shown in tables~\ref{tab:rc_scaling_relation}, \ref{tab:deltac_scaling_relation}, and \ref{tab:Mc_scaling_relation}. These values correspond to the fits shown in figures~\ref{fig:simulation_fitted_parameters} and \ref{fig:Rc_3kpc_ic_comparisons} obtained using Theil-Sen regression, including the results for the SPARC dataset \citep{Li2020} and the Milky Way dSphs \citep{Salucci2012}.

\bgroup 
\def\arraystretch{1.3}
\begin{table}
\caption{The fitted scaling parameters for $r_c(M_{200})$, eq. \eqref{eq:rc_fitting_func}, at various redshifts. The results from the SPARC dataset and the Milky Way dSphs are included under the label "Data". \newline}              
\label{tab:rc_scaling_relation}      
\centering                                      
\begin{tabular}{l | c c c c c c c | c c c c c c c }           
\hline\hline
& \multicolumn{7}{c|}{$r_{c,10}$ [$\text{kpc}$]} & \multicolumn{7}{c}{$\alpha$}  \\
\hline
z   & $3$ & $2.5$ & $2$ & $1.5$ & $1$ & $0.75$ & $0.5$ & $3$ & $2.5$ & $2$ & $1.5$ & $1$ & $0.75$ & $0.5$ \\
\hline
Rc1   & 1.2 & 1.3 & 1.2 & 1.5 & 1.6 & 1.4 & 1.4 & 0.091 & 0.095 & 0.070 & 0.091 & 0.10 & 0.064 & 0.058 \\
Rc3   & 2.4 & 2.5 & 2.8 & 3.0 & 3.2 & 3.2 & 3.6 & 0.084 & 0.080 & 0.096 & 0.080 & 0.080 & 0.038 & 0.062 \\
Rc3-b & 2.7 & 2.9 & 3.4 & 3.8 & 4.2 & 4.3 & 3.5 & 0.17   & 0.17 & 0.16 & 0.14 & 0.11 & 0.11 & 0.29 \\
Rc3-c & 2.5 & 2.6 & 2.8 & 3.1 & 3.2 & 3.3 & 3.6 & 0.10  & 0.073 & 0.086 & 0.091 & 0.072 & 0.052 & 0.056 \\
Rc3-d & 2.6 & 2.8 & 3.1 & 3.3 & 3.5 & 3.6 & 3.7 & 0.096 & 0.080 & 0.11 & 0.10 & 0.082 & 0.080 & 0.044 \\
Rc10  & 5.5 & 6.0 & 6.6 & 7.2 & 8.3 & 8.5 & 9.0 & 0.081 & 0.081 & 0.097 & 0.092 & 0.079 & 0.084 & 0.080 \\
Data  & \multicolumn{7}{c|}{0.86} & \multicolumn{7}{c}{0.45} \\
\hline\hline
\end{tabular}
\end{table}
\egroup 

\bgroup 
\def\arraystretch{1.3}
\begin{table}
\caption{The fitted scaling parameters for $\delta_c(M_{200})$, eq. \eqref{eq:deltac_fitting_func}, at various redshifts. The results from the SPARC dataset and the Milky Way dSphs are included under the label "Data". \newline}             
\label{tab:deltac_scaling_relation}      
\centering                                      
\begin{tabular}{l | c c c c c c c | c c c c c c c }          
\hline\hline
& \multicolumn{7}{c|}{$\delta_{c,10} \times 10^3$} & \multicolumn{7}{c}{$\beta$} \\
\hline
z   & $3$ & $2.5$ & $2$ & $1.5$ & $1$ & $0.75$ & $0.5$ & $3$ & $2.5$ & $2$ & $1.5$ & $1$ & $0.75$ & $0.5$ \\
\hline
Rc1   & 11 & 13 & 20 & 21 & 30 & 41 & 72 & 0.44 & 0.45 & 0.49 & 0.47 & 0.46 & 0.47 & 0.57 \\
Rc3   & 3.2 & 3.6 & 4.0 & 5.2 & 7.2 & 9.5 & 11  & 0.45 & 0.45 & 0.43 & 0.47 & 0.49 & 0.58 & 0.56 \\
Rc3-b & 2.5 & 2.7 & 2.9 & 3.3 & 4.2 & 5.2 & 3.8 & 0.29 & 0.31 & 0.33 & 0.35 & 0.44 & 0.44 & 0.97 \\
Rc3-c & 2.7 & 3.5 & 3.9 & 4.9 & 6.9 & 8.5 & 11  & 0.41 & 0.47 & 0.45 & 0.44 & 0.49 & 0.54 & 0.57 \\
Rc3-d & 2.7 & 3.1 & 3.4 & 4.2 & 6.0 & 7.6 & 9.6 & 0.43 & 0.45 & 0.42 & 0.43 & 0.49 & 0.5 & 0.59 \\
Rc10  & 0.67 & 0.73 & 0.83 & 1.0 & 1.2 & 1.4 & 1.6 & 0.44 & 0.45 & 0.44 & 0.44 & 0.48 & 0.48 & 0.51 \\
Data  & \multicolumn{7}{c|}{630} & \multicolumn{7}{c}{-0.33} \\
\hline\hline
\end{tabular}
\end{table}
\egroup 

\bgroup 
\def\arraystretch{1.3}
\begin{table}
\caption{The fitted scaling parameters for $M_c(M_{200})$, eq. \eqref{eq:Mc_fitting_func}, at various redshifts. The results from the SPARC dataset and the Milky Way dSphs are included under the label "Data". \newline}             
\label{tab:Mc_scaling_relation}      
\centering                                      
\begin{tabular}{l | c c c c c c c | c c c c c c c }          
\hline\hline
& \multicolumn{7}{c|}{$M_{c,10}$ [$10^{8}M_{\odot}$]} & \multicolumn{7}{c}{$\gamma$} \\
\hline
z   & $3$ & $2.5$ & $2$ & $1.5$ & $1$ & $0.75$ & $0.5$ & $3$ & $2.5$ & $2$ & $1.5$ & $1$ & $0.75$ & $0.5$ \\
\hline
Rc1   & 4.4 & 4.5 & 3.5 & 3.6 & 3.5 & 2.7 & 2.7 & 0.72 & 0.74 & 0.70 & 0.74 & 0.78 & 0.75 & 0.76 \\
Rc3   & 9.8 & 8.8 & 8.5 & 7.5 & 6.5 & 6.0 & 5.8 & 0.71 & 0.69 & 0.72 & 0.71 & 0.73 & 0.71 & 0.75 \\
Rc3-b & 11 & 10 & 10 & 9.7 & 8.5 & 7.9 & 4.7 & 0.81 & 0.82 & 0.80 & 0.78 & 0.77 & 0.77 & 0.97 \\
Rc3-c & 10 & 8.9 & 8.4 & 7.7 & 6.6 & 6.2 & 5.8  & 0.72 & 0.69 & 0.71 & 0.72 & 0.72 & 0.72 & 0.75 \\
Rc3-d & 10 & 9.6 & 9.2 & 8.3 & 7.0 & 6.7 & 5.8  & 0.71 & 0.69 & 0.74 & 0.73 & 0.74 & 0.75 & 0.73 \\
Rc10  & 25 & 24 & 23 & 20 & 18 & 16 & 14 & 0.69 & 0.70 & 0.73 & 0.72 & 0.72 & 0.74 & 0.75 \\
Data  & \multicolumn{7}{c|}{1.2} & \multicolumn{7}{c}{1.1} \\
\hline\hline
\end{tabular}
\end{table}
\egroup 
\twocolumn

\end{document}